\documentclass[letterpaper,twocolumn,10pt]{article}
\usepackage{usenix}
\usepackage[ruled,vlined,linesnumbered]{algorithm2e}
\usepackage{amsmath,amsfonts}
\usepackage{graphicx}
\usepackage{textcomp}
\usepackage{xcolor}
\usepackage{tikz}

\usepackage{xurl}
\usepackage[ruled,vlined,linesnumbered]{algorithm2e}
\usepackage{comment}
\usepackage{soul, color}
\usepackage{subcaption}
\usepackage{threeparttable}
\usepackage{multirow}
\usepackage{makecell}
\usepackage{booktabs}
\usepackage{verbatim}
\usepackage{epstopdf}
\usepackage{rotating}
\usepackage{listings}
\usepackage{arydshln}
\usepackage{listing}
\usepackage{paralist}
\usepackage{arydshln}
\usepackage{algpseudocode}
\usepackage{endnotes}
\usepackage{xspace}
\usepackage{pifont}
\usepackage{wasysym}
\usepackage{amsthm}
\usepackage{extdash}
\usepackage{tabularx}
\usepackage{float}
\usepackage{etoolbox}
\usepackage{enumitem}
\usepackage{pdfpages}
\usepackage{todonotes}
\usepackage{array}    % For table wrapping and more complex table structures
\usepackage{adjustbox}
\usepackage{hyperref}
\usepackage{cite}
\usepackage{xcolor}

\definecolor{codeblue}{RGB}{0, 0, 180}
\definecolor{codegreen}{RGB}{0, 128, 0}
\definecolor{codered}{RGB}{170, 34, 255}
\definecolor{codegray}{RGB}{120, 120, 120}

\lstdefinestyle{C_style}{
  language=C,
    belowcaptionskip=1\baselineskip,
    breaklines=true,
    frame=none,
    numbers=none,
    keepspaces = true,
    tabsize=2,
    basicstyle=\scriptsize\ttfamily,
    keywordstyle=\color{black!40!black},
    commentstyle=\itshape\color{purple!40!black},
    frame=tb, % <-- This line adds top and bottom lines
	moredelim=**[is][\bfseries\itshape\color{green!70!black}]{`}{`},
  moredelim=**[is][\bfseries\color{red!70!black}]{$}{$}
}

\lstdefinestyle{customc}{
  language=C,
  basicstyle=\ttfamily\tiny,
  keywordstyle=\color{blue},
  commentstyle=\color{green!40!black},
  stringstyle=\color{orange},
  numbers=left,
  numberstyle=\tiny,
  numbersep=5pt,
  frame=single,
  breaklines=true,
  breakatwhitespace=true,
  tabsize=4,
  morekeywords={int, char, if, else, while, return}
}
\usetikzlibrary{snakes,arrows,shapes}

\newcommand{\pushan}{\textsc{Pushan}\xspace}
\newcommand{\system}{\pushan{}\xspace}

\newcommand{\code}[1]{\texttt{#1}}

\renewcommand{\paragraph}[1]{\smallskip \noindent\textbf{#1.}\hspace{1em}}

\definecolor{graph-blue}{RGB}{59, 117, 175}
\definecolor{graph-yellow}{RGB}{242, 169, 59}

\definecolor{pinegreen}{rgb}{0.0, 0.47, 0.44}
\definecolor{rossocorsa}{rgb}{0.83, 0.0, 0.0}
\newcommand{\cmark}{\textcolor{pinegreen}{\ding{51}}}
\newcommand{\xmark}{\textcolor{rossocorsa}{\ding{56}}}

\newtoggle{diff}
\togglefalse{diff}

\hypersetup{
    colorlinks,
    linkcolor={red!80!black},
    citecolor={blue},
    urlcolor={blue!80!black}
}

\begin{document}
\date{} % don’t print date

\title{\Large \bf Pushan: Trace-Free Deobfuscation of Virtualization-Obfuscated Binaries}

% \author{
% {\rm Blinded for submission}\\
% Affiliation withheld
% }

% \author{\IEEEauthorblockN{Michael Shell}
% 	\IEEEauthorblockA{Georgia Institute of Technology\\
% 		someemail@somedomain.com}
% 	\and
% 	\IEEEauthorblockN{Homer Simpson}
% 	\IEEEauthorblockA{Twentieth Century Fox\\
% 		homer@thesimpsons.com}
% 	\and
% 	\IEEEauthorblockN{James Kirk\\ and Montgomery Scott}
% 	\IEEEauthorblockA{Starfleet Academy\\
% 		someemail@somedomain.com}}


%Telephone: (800) 555--1212, Fax: (888) 555--1212}

\author{
{\rm
Ashwin Sudhir,
Zion Leonahenahe Basque,
Wil Gibbs,
Ati Priya Bajaj,
Pulkit Singh Singaria,
}\\
{\rm
Mitchell Zakocs,
Jie Hu,
Moritz Schloegel,
Tiffany Bao,
Adam Doupe,
}\\
{\rm
Yan Shoshitaishvili,
Ruoyu Wang
}\\
\textit{Arizona State University}\\
\textit{CISPA Helmholtz Center for Information Security}\\
\textit{\{asudhir1,zbasque,wfgibbs,atipriya,psingari,mzakocs,jiehu12,tbao,doupe,yans,fishw\}@asu.edu}\\
\textit{moritz.schloegel@cispa.de}
}


\maketitle

\begin{abstract}
In the ever-evolving battle against malware, binary obfuscation techniques are a formidable barrier to the effective analysis of malware by both human security analysts and automated systems. In particular, virtualization or VM-based obfuscation is one of the strongest protection mechanisms that evade automated analysis. Despite widespread use of virtualization in practice, existing automated deobfuscation techniques suffer from three major drawbacks. First, they only work on execution traces, which prevents them from recovering \emph{all} logic in an obfuscated binary. Second, they depend on dynamic symbolic execution, which is expensive and does not scale in practice. Third, they cannot generate ``well-formed'' code, which prevents existing binary decompilers from generating human-friendly decompilation output.

This paper introduces \system, a novel and generic technique for deobfuscating virtualization-obfuscated binaries while overcoming the limitations of existing techniques. 
\system is trace-free and avoids path-constraint accumulation by using VPC-sensitive, constraint-free symbolic emulation to recover a complete CFG of the virtualized function. 
It is the first approach that also decompiles the protected code into high-quality C pseudocode to enable effective analysis. 
Crucially, \system circumvents the traditional reliance on path satisfiability, which is a known NP-hard problem that hampers the scalability of existing deobfuscation techniques. 
We evaluate \system on a diverse set of more than $1,000$ binaries, including targets protected by academic state of the art, Tigress, and by the commercial-strength obfuscators VMProtect and Themida. 
\system successfully deobfuscates these binaries, retrieves their \emph{complete} CFGs, and decompiles them to C pseudocode. We further demonstrate practical applicability by analyzing a previously unanalyzed VMProtect-obfuscated malware sample from VirusTotal, where our decompiled output enables LLM-assisted code simplification, reuse and program understanding, making our approach the first to enable effective end-to-end analysis of code protected by virtualization.

% that is semantically equivalent to the original unobfuscated programs. % It even enables code reuse through recompilation.
% We further commit to open science by pledging to open source \system and our evaluation artifacts, fostering further research and development in the field of automated VM-deobfuscation and malware analysis.

\end{abstract}


% \fi

% CONTENT BODY
%-------------------------------------------------------------------------------
\section{Introduction}%
\label{s:intro}
%-------------------------------------------------------------------------------

% Malware is a big problem, and VM-based obfuscation techniques make them difficult to analyze

Over 50 years have passed since the birth of the first computer virus in 1971, and malware is still a critical and widespread threat to Internet users.
Security companies report that they detected more than 500,000 unique malware samples per day in 2025, which is a 7\% increase over 2024~\cite{kaspersky_detection_2025}.
The constant flow of newly discovered malware has driven the recruitment of human security analysts and the development of additional intelligent automated techniques for malware analysis.
To thwart such attempts, malware authors turned to protection schemes, with virtualization-based binary obfuscation techniques (also referred to as Virtual Machine (VM)-based obfuscation) being the most notable~\cite{assis2019comparative}.

A virtualization-based obfuscator protects a binary program from both manual and automated reverse engineering by translating the original instructions into virtual machine (VM) bytecode that uses a custom, usually randomized instruction set.
Then, it injects one or more VM interpreters (also called virtual CPUs) into the protected binary, which interpret the newly generated VM bytecode during run time.
Essentially, virtualization-based obfuscation hides the original control flow, which is often critical for reverse engineering~\cite{remind}, and instead exposes only the control flow of the VM interpreters.
To reverse engineer such a binary, security analysts must first reverse engineer the VM interpreters, which is tedious, time-consuming, and error-prone.

% Existing tools and techniques suck

% While there are existing tools that attempt to automate and generalize this process, we we still lack output that can actually help reverse engineers easily understand the control flow and the semantics of the original program.
% In this paper we aim to address these two problems by providing a complete control flow graph and a C like pseduo code that is easy to understand.

% While researchers have proposed many techniques for automatically deobfuscating virtualization-obfuscated binaries~\cite{yadegari2015generic, salwan2018symbolic, xu2018vmhunt,bardin2017backward}, three limitations prevent the adoption of academic state of the art.

While academic efforts to automate the deobfuscation of virtualization-obfuscated binaries are extensive~\cite{yadegari2015generic, salwan2018symbolic, xu2018vmhunt, bardin2017backward, blazytko2017syntia, menguy2021xyntia}, three key limitations hinder their practical adoption:
(1) \textit{Incomplete Path Coverage}: Existing techniques are primarily trace-based, reasoning about one execution path at a time to recover a control flow graph (CFG).
Given the complex logic and multitude of paths in real-world malware, achieving complete path coverage is virtually impossible, potentially leaving analysts with an incomplete CFG and missing critical logic.
(2) \textit{Low Scalability}: These techniques rely on dynamic symbolic execution~\cite{balliu2012encover} for path exploration, which is computationally expensive and inherently unscalable due to the path explosion problem and the NP-completeness~\cite{shoshitaishvili2016sok} of satisfiability checking.
Furthermore, dynamic symbolic exploration is notoriously brittle and easily defeated by common obfuscation countermeasures~\cite{seto2019preventing,wang2011linear,ollivier2019kill,schloegel2022loki,banescu2016code}.
(3) \textit{Low Usability}: Existing techniques output raw instruction traces that are not conducive to downstream analysis.
Security analysts rely on familiar tools like binary decompilers to interpret code~\cite{yakdan2016dreampp,botacin2019revenge,votipka2021investigation,basque2026decompiling}, which improve the speed and accuracy of malware analysis~\cite{votipka2020observational}.
However, these tools expect structured, ``well-formed'' assembly code and a complete CFG~\cite{behner2025sok} for correct performance.
As Section~\ref{sec:motivation} shows, even industry-standard tools like the Hex-Rays Decompiler fail to produce clean C pseudocode from deobfuscated traces, forcing analysts back into tedious manual reversing.

In this paper, we present \system, a scalable technique that deobfuscates virtualization-obfuscated binaries in a \emph{trace-free} manner, recovers a \emph{complete} CFG of the deobfuscated program, and generates decompiled C pseudocode as the output.
\system builds on the concept of Virtual Program Counter (VPC) sensitivity~\cite{kinder2012towards}, which is drawn from the concept of sensitivities in program analysis:
We can recover the \emph{original} CFG of a virtualization-obfuscated binary by making CFG recovery sensitive to both the actual program counter (which always points to instructions in an VM interpreter) and the VPC, which always points to a location in the VM bytecode region.
To achieve VPC sensitivity, we design a new CFG recovery algorithm based on \emph{constraint-free symbolic emulation}, which symbolically evaluates instructions without accumulating path constraints or performing path exploration.
\system recovers a VPC-sensitive, flattened CFG (termed a \emph{flat CFG}) that contains both the (now flattened) interpreter logic and the logic of the original program.
It then iteratively performs semantics-preserving simplifications to simplify the flat CFG.
The goal of simplification is to remove any logic related to VM interpreters and only retain the logic of the original, unobfuscated code.
Lastly, \system uses a custom binary decompiler to generate C pseudocode, which makes malware easier to analyze~\cite{votipka2020observational}.
This advantage largely stems from the correctness of the generated CFG used during decompilation~\cite{dewolf, behner2025sok}.

We evaluate \system on several datasets to understand its correctness, capabilities, and impact.
First, we used 14 programs, including malware samples, open-source software, and synthetic binaries, obfuscated using the widely adopted, commercial obfuscators VMProtect~\cite{vmprotect} and Themida~\cite{themida}, yielding 28 unique obfuscated binaries.
Our evaluation shows that \system is not hindered by the path satisfiability problem, can fully recover the CFGs, and achieves 100\% similarity to the original CFGs for 17 out of the 28 cases, with high similarity in the remaining samples.
Then, we evaluated \system on 1,000 binaries generated with Tigress using three different VM configurations, finding that \system fully analyzed and successfully deobfuscated and decompiled 988 of them.
We compared this against the state of the art~\cite{salwan2018symbolic}, which only recovered 68 complete CFGs.
We also evaluated \system on five CTF challenges obfuscated with bespoke VM implementations.
It succeeds in all five samples and for one even revealed the embedded flag value in the decompiled output, highlighting its ability to recover semantically meaningful logic from custom VM designs.

% Critically, we demonstrate that the decompilation output generated by \system facilitates code reuse, such as enabling protocol interoperability, by allowing the recompilation of an obfuscated malware function from the \emph{blaster} sample, where we needed to change only 53 lines out of 146, all of which were minor modifications.
% This demonstrates that \system produces decompiler-consumable control flow that enables practical decompilation of virtualization-obfuscated binaries.

Finally, we evaluated \system on a real-world VMProtect-obfuscated binary sample and recovered the logic of a virtualization-obfuscated function in the binary.
This allowed us to determine the true nature of this binary and demonstrates that \system produces decompiler-consumable control flow that enables practical decompilation of virtualization-obfuscated binaries.
To our best knowledge, while this sample has existed since 2018 (according to VirusTotal), this is the first thorough analysis of it.
%approach
% In our work we focus only on the virtualization aspect of obfuscation and we believe that our work can be integrated with other existing methods such as those for MBA de-obfuscation~\cite{liu2021mbablast} to build an end to end solution for complete virtual machine deobfuscation.

%analysable c pesudo code; 2 diagrams- pipeline overview and result overview

\smallskip
\noindent
\textbf{\textit{Contributions.}}
This paper makes three key contributions:

\begin{itemize}[nosep]
    \item
        We review state-of-the-art devirtualization approaches and identify a fundamental problem that limits their scalability and usability.
    \item
        We propose \system, a scalable deobfuscation framework based on \emph{VPC-sensitive, constraint-free symbolic emulation} that recovers complete control flow from virtualization-obfuscated binaries.
        %\system recovers a complete CFG of the original binary and generates high-quality C pseudocode of the recovered binary.
    \item
        We evaluate \system on binaries protected by commercial-strength obfuscators and demonstrate its ability to recover complete CFGs and produce decompiled code semantically consistent with original code.
    %Evaluating \system on binaries protected by commercial-strength obfuscators, we find it successfully deobfuscates them, generates complete CFGs, and produces decompiled C pseudocode that is semantically consistent with the original programs.
\end{itemize}

% \noindent
% We will release the source code of \system as well as our evaluation artifacts upon the publication of this paper.

%-------------------------------------------------------------------------------
% \section{Background}%
\section{Virtual Machine Deobfuscation}%
\label{s:background}%
\label{sec:background:vmdeobfuscation}

% We briefly introduce relevant background for our paper.

\begin{table*}[tbp]
  \centering
  \caption{A qualitative comparison of \system against existing techniques for virtualization-based deobfuscation.}%
  \label{tab:system_comparison}
  {\footnotesize
  \begin{tabular}{lllccc}
    \toprule
    & Key Techniques & Output & Complex Programs & CFG Complete? & Code Decompilable? \\
    \midrule
    Kinder~\cite{kinder2012towards}          & abstract interpretation, VPC sensitivity & calls+arguments & ?\rlap{$^1$} & -- & \xmark \\
    Coogan~\cite{coogan2011deobfuscation} & tracing, slicing & subtrace & \cmark & -- & \xmark \\
    Yadegari~\cite{yadegari2015generic}   & tracing, taint analysis & CFG & \cmark & \xmark & \xmark \\
    \textsc{VMHunt}~\cite{xu2018vmhunt} & tracing, symbolic execution & Trace & \cmark & -- & \xmark \\
    Salwan~\cite{salwan2018symbolic} & tracing, taint, formula slicing & CFG & \xmark\rlap{$^2$} & \xmark & \cmark \\
    \system & emulation, VPC sensitivity & CFG & \cmark & \cmark & \cmark \\
    \bottomrule
    \multicolumn{6}{l}{\footnotesize{}$^1$: Only one toy example tested \hspace{1cm} $^2$: Supports obfuscated code that contains not more than one or two branches} \\
  \end{tabular}
  }
% \end{adjustbox}
\end{table*}

Many obfuscation techniques exist, including Mixed Boolean-Arithmetic (MBAs)~\cite{zhou2007information} or Opaque Predicates~\cite{collberg1997taxonomy,collberg1998manufacturing}.
One of the most powerful techniques is \emph{virtualization-based obfuscation}, also referred to as \emph{VM-based obfuscation}.
It runs the to-be-protected code inside a custom interpreter that uses a custom bytecode representation.
% An example of how this interpreter can look like is provided in Figure~\ref{fig:virtualization_example}.
% The presented example features a centralized fetch-decode-execute loop that interprets the VM bytecode and invokes the respective VM handler.
Figure~\ref{fig:virtualization_example} illustrates a typical interpreter design, featuring a centralized fetch-decode-execute loop that dispatches VM bytecode instructions to corresponding handlers.
Different architectures are possible, including threaded code~\cite{dewar1975indirect,piumarta1998optimizing}, where the address of the next bytecode (direct threading) or the fetch and decode step (indirect threading) are inlined into the VM handlers, avoiding the easy to identify VM dispatcher loop.
The Virtual Program Counter (VPC), akin to its analogous counterpart on a CPU, tracks the execution through the bytecode.

\definecolor{handlercolor}{HTML}{0070c0}
\definecolor{bytecodecolor}{HTML}{4ea72e}

\begin{figure}[tb]
    \centering
    \includegraphics[width=0.8\columnwidth]{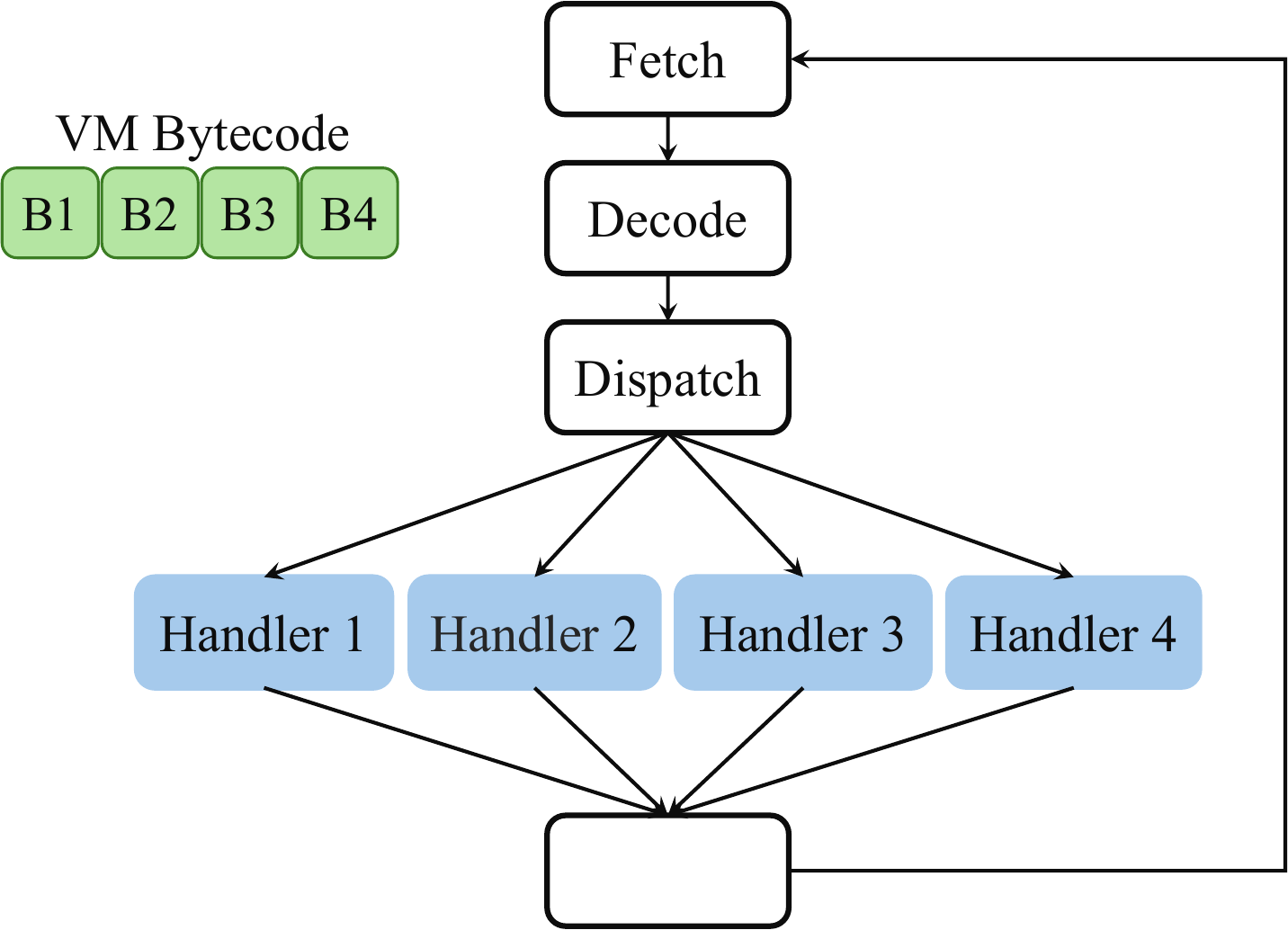}
    \caption{An illustration of virtualization-based obfuscation, where the to-be-protected code has been translated into \textcolor{bytecodecolor}{\textbf{bytecode}}.
    A \emph{fetch-decode-execute} loop with a central dispatcher dispatches execution to individual \textcolor{handlercolor}{\textbf{VM handler}}s.
    Each bytecode instruction will be handled by a VM handler.
    %% Long form
    % Example of VM-based obfuscation. The original code has been translated to a new ISA and is placed in the program as \textcolor{green}{bytecode}. The program is then bundled with the depicted interpreter, which can execute this bytecode. It typically follows a \emph{fetch-decode-execute} loop and often features a central \emph{dispatcher}. Also noteworthy is the wide number of \textcolor{blue}{handlers}.%
    } %
    \label{fig:virtualization_example}
\end{figure}

%% END NEW BACKGROUND INTRO

% \subsection{VM Deobfuscation}%
% \label{sec:background:vmdeobfuscation}

Removing VM-based obfuscation requires (1) identifying the VM, (2) mapping VM handlers, (3) reconstructing handler semantics, and (4) developing a disassembler for the VM.
Only after these steps can an analyst begin studying the underlying program logic.
To alleviate this burden, various works have proposed automating Steps (2) to (4).

Table~\ref{tab:system_comparison} lists the state-of-the-art solutions in VM deobfuscation with the goal of recovering the original code.
Studying their underlying primitives, we can identify two philosophies: techniques that rely on tracing and mount an analysis on these traces, and Kinder's VPC-sensitive abstract interpretation.

\smallskip
\noindent
\textbf{Tracing-based analysis.}
Trace-based techniques bypass virtualization by observing VM behavior on concrete executions.
% Fundamentally, they (1) trace the program execution for one input.
% Then, (2) they can extract information from this trace or simplify it, for example, by using slicing or taint analysis to identify instructions not depending on the user input.
They (1) capture an execution trace for a given input, and (2) simplify the trace using techniques such as taint analysis to isolate the underlying logic from input-dependent or obfuscation-related noises.
Some techniques use program synthesis to find equivalent but simpler expressions for VM handlers~\cite{blazytko2017syntia,menguy2021xyntia}.
While they differ in implementation details and complexity, this approach is fundamentally limited to one execution path and cannot capture the \emph{complete} program behavior.
Optionally, (3) they may derive input cases (e.g., by querying an SMT solver) to generate additional exeuction traces, which is popular among prior work~\cite{coogan2011deobfuscation,yadegari2015generic,xu2018vmhunt,salwan2018symbolic}.
% Yadegari~\etal~\cite{yadegari2015generic} and Salwan~\etal~\cite{salwan2018symbolic} mitigate this by employing symbolic execution to generate new inputs, which are then used to generate additional execution traces.
However, symbolic execution is particularly brittle in the context of obfuscation, which often employs safeguards~\cite{banescu2016code, ollivier2019kill}, and it still does not guarantee complete path coverage, as we will show in the following section.

\smallskip
\noindent
\textbf{VPC-sensitive abstract interpretation.}
Kinder~\cite{kinder2012towards} proposes VPC-sensitive abstract interpretation.
While traditional static analysis would merely analyze the VM's interpreter, VPC sensitivity accounts for this layer of indirection. The key insight is that in programs under virtualization-based obfuscation, different abstract states exist for different execution contexts, but are merged under the same program location.
For example, the VM's \texttt{add} handler may be called from different contexts, whenever addition is needed.
We can account for this by adding a dimension, bytecode location sensitivity, to our analysis~\cite{kinder2012towards}:
By tracking the location of the bytecode (i.e., point in program execution) calling the VM handler, static analysis can now differentiate between calls to the same location.
Unfortunately, prior work does not scale; It only showed this technique on a toy program to extract specific system calls and their arguments.

\medskip
Existing techniques either remain coverage-limited when relying on traces, or provide partial analysis results rather than fully deobfuscated program representations.
Scalable recovery of complete CFGs and human-consumable output for binaries protected by commercial-strength virtualizers remains unsolved in practice.

\section{Challenges}%
\label{sec:motivation}

\begin{figure}[h!]
    \centering
    % Row 1
    \begin{subfigure}[b]{0.58\linewidth}
\begin{lstlisting}[
            language=C,
            basicstyle=\ttfamily\scriptsize,
            numbers=left,
            numbersep=3pt,
            frame=lines,
            breaklines=true,
            breakindent=1em
        ]
while (input_num != 0) {
    hash = ((hash << 5) + hash) + (input_num % 10);
    hash = hash ^ (hash >> 3);
    input_num /= 10;
}
hash = hash & 0xFFFFFFFFFFFFFFFF;

// the hash of 4294967295
if (hash == 226292709949525)
    puts("Correct!");
else
    puts("Wrong!");
\end{lstlisting}
        % \vspace{-0.8em}
        \caption{Original source code}
        \label{lst:pre_image_hash}
    \end{subfigure}%
    \hfill
    \begin{subfigure}[b]{0.42\linewidth}
        \centering
        \includegraphics[width=\textwidth]{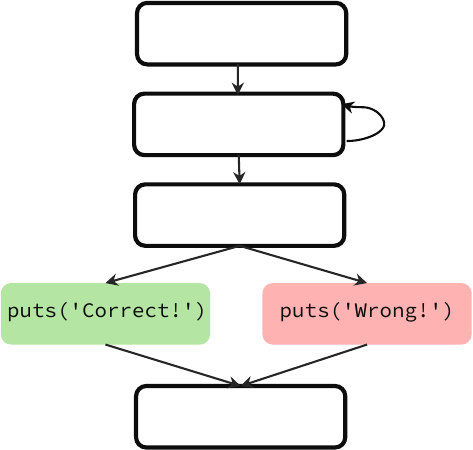}
        \caption{\textbf{Unobfuscated} CFG}
        \label{fig:hash_pushan_orig_graph}
    \end{subfigure}%
    \caption{Motivating example.}
\end{figure}

\begin{figure}[h!]
    \centering
    \begin{subfigure}[b]{0.28\linewidth}
        \centering
        \includegraphics[height=4cm, keepaspectratio]{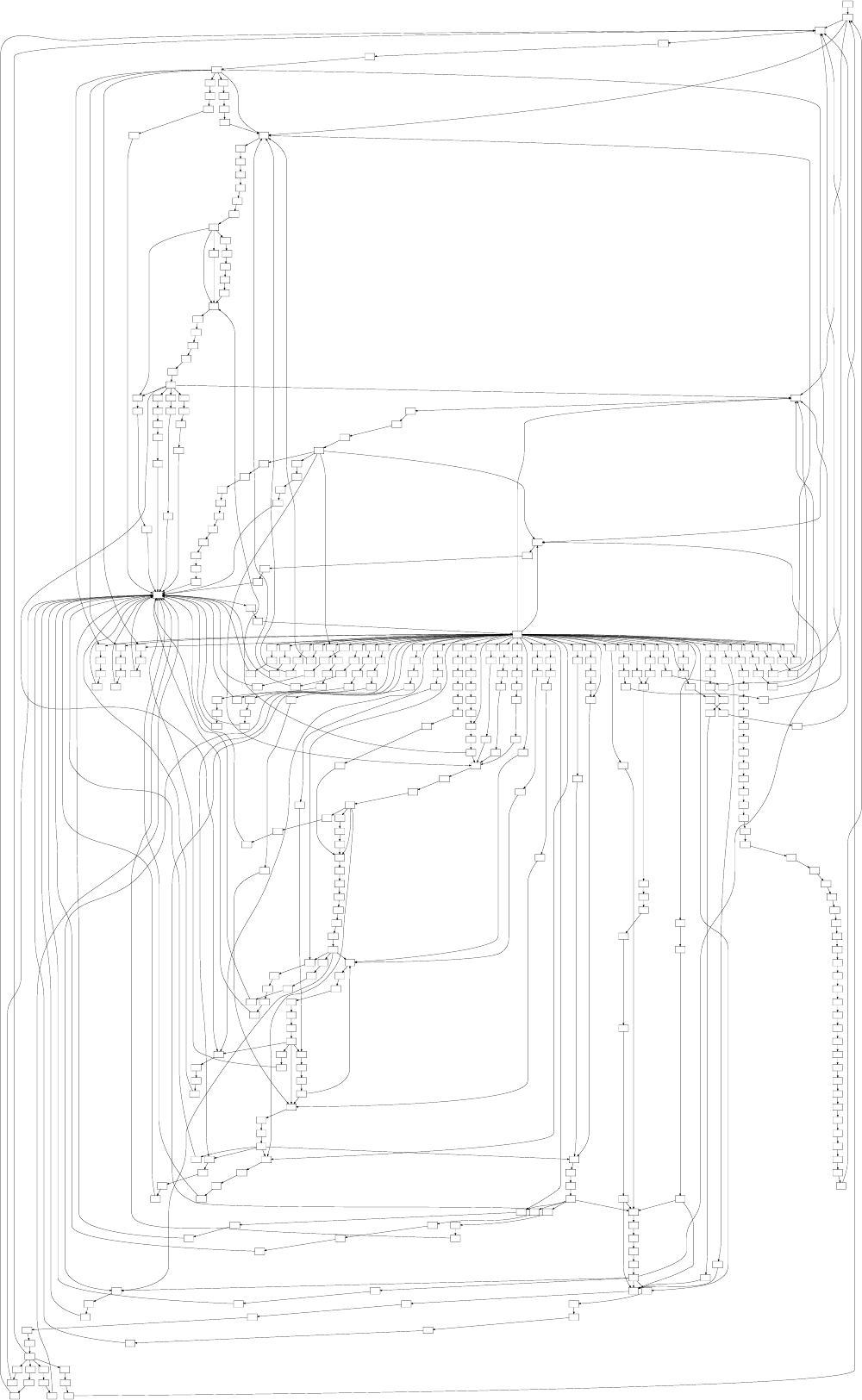}
        \caption{}
        % \caption{VM-obfuscated CFG}
        \label{fig:hash_vm_graph}
    \end{subfigure}
    \hfill
    \begin{subfigure}[b]{0.24\linewidth}
        \centering
        \includegraphics[width=0.8\textwidth]{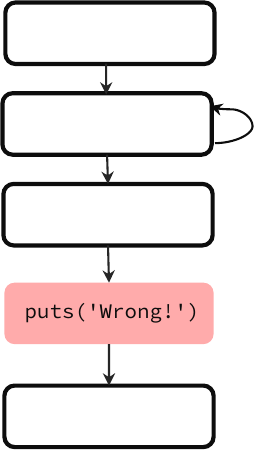}
        \caption{}
        % \caption{\textbf{Incomplete} deobfuscated CFG (trace-based approaches)}
        \label{fig:hash_trace_graph}
    \end{subfigure}%
    \hfill
    \begin{subfigure}[b]{0.40\linewidth}
        \centering
        \includegraphics[width=\textwidth]{figures/hash_orig_cfg_cropped.pdf}
        \caption{}
        % \caption{Deobfuscated \textbf{complete} CFG (\system)}
        \label{fig:hash_pushan_graph}
    \end{subfigure}
    \caption{CFGs recovered (a) directly from VM-obfuscated code, (b) by trace-based approach, and (c) by \system.}
\end{figure}

Two challenges limit prior work and motivate our approach.
Consider the example program in Figure~\ref{lst:pre_image_hash}, which implements a simple hash-based input checker.
While the original CFG (Figure~\ref{fig:hash_pushan_orig_graph}) is straightforward, the VMProtect-obfuscated version (Figure~\ref{fig:hash_vm_graph}) is significantly larger and structurally unrecognizable.
This highlights the main obstacle:
Conventional CFG recovery yields the structure of the VM interpreter rather than the original control flow of the protected program.

\smallskip
\noindent
\textbf{Challenge 1: Incomplete program behaviors captured.}
Prior work on VM deobfuscation relies on program \emph{traces} to identify executed code.
% While this simplifies analysis, it limits the program behavior captured to the one observed:
A single trace, as in VMHunt~\cite{xu2018vmhunt} and Coogan~et~al.~\cite{coogan2011deobfuscation}, is insufficient to retrieve the \emph{complete} behavior.
Even the simple \texttt{hash} example requires at least two traces to capture both the ``success'' and ``wrong'' paths.
A single trace (assuming it triggers the ``wrong'' path) will lead to an incomplete deobfuscated CFG as shown in Figure~\ref{fig:hash_trace_graph}.

% OLD:
% Having a single trace, as VMHunt and Coogan~et~al.~\cite{coogan2011deobfuscation} use, is not sufficient to recover the full behavior of the obfuscated code. Only behavior that has been executed during the trace will be uncovered. This challenge does not apply to Kinder, who uses a static approach, and while Salwan and Yadegari use symbolic or concolic execution of the trace to derive new inputs for every observed branch, this has its own challenges. As Table~\ref{tab:system_comparison} shows, only three approaches can restore a CFG, and only \pushan succeeds in retrieving the \emph{complete} CFG.

\smallskip
\noindent
\textbf{Challenge 2: Constraint solving-based input generation does not scale.}
Two works attempt to address Challenge~1~\cite{salwan2018symbolic,yadegari2015generic}.
To observe \emph{all feasible} input paths, they turn to repeated dynamic symbolic execution (DSE).
They first symbolically execute a trace and collect path constraints, then negate these constraints at branching points and generate an input to trigger the alternate branch.
They repeat the process to discover more traces, hoping to reach full path coverage of the protected code.
% This input-generation loop is intended to improve coverage, but its scalability hinges on solving the path constraints induced by the protected code.

Unfortunately, this approach inherits the downsides of symbolic execution:
It suffers from \emph{path explosion}, where there are (nearly) an infinite number of paths in complex, real-world programs, and it lacks a model for the \emph{environment}, which complicates its application in general.
Most importantly, both obfuscation and the original program logic can field \emph{constraints that are difficult to solve} for theorem provers~\cite{banescu2016code,schloegel2022loki}.
The sample program \texttt{hash} will only execute \texttt{puts("Correct!")} if the user input hashes to a certain value.
Without having the intended user input (in this case, \texttt{4294967295}), a theorem prover must break the hash algorithm's preimage resistance and find an input that satisfies \texttt{hash == 226292709949525}.
% To simulate the constraint solving required by DSE-based approaches to trigger the \texttt{puts("Correct")} branch
We attempted to solve for the correct input using angr and Z3 on the unobfuscated binary.
Z3 was unable to find a valid 64-bit input within three hours, which shows the practical limitations of constraint-solving-based approaches.
These weaknesses hamper the scalability of symbolic execution as a means of generating new inputs to discover more program behaviors.
Obfuscation has moved to specifically target symbolic execution~\cite{schloegel2022loki,ollivier2019kill,wang2011linear,xu2018manufacturing}, rendering it even more ineffective.
% Unfortunately, both approaches~\cite{salwan2018symbolic,yadegari2015generic} that use multiple traces are impacted by this.

\section{The Design of \system}%
\label{sec:design}

% To address the limitations identified in Section~\ref{sec:motivation}, we build \system around two key insights.
% As discussed in Challenge~1, trace-based analyses fail to recover complete program behavior, while Challenge~2 shows that symbolic-execution–based input generation does not scale due to path explosion and hard-to-solve constraints.
%
\system aims to recover a \emph{complete control-flow structure} of a virtualization-obfuscated binary in a trace-free manner.
Our solution depends on two insights:
% We therefore design an analysis that reasons directly about control-flow reachability and instruction semantics, without relying on execution traces or path satisfiability.
% The following two insights explain how \system avoids both path explosion and constraint-solving bottlenecks while recovering complete CFGs from virtualization-obfuscated binaries.

\smallskip
\noindent\textbf{Bounding symbolic state growth during CFG construction.}
\system recovers the CFG by \emph{symbolically emulating} the program from the entry point and incrementally constructing the CFG.
Each time symbolic emulation encounters a control-flow transfer, a corresponding node and edge are added to the graph.
Traditional DSE must preserve a distinct symbolic state for every execution path reaching a program location, since path feasibility depends on how that location was reached.
This leads to path explosion due to branches and loops.

In contrast, we observe that CFG recovery is a \emph{structural} problem. % instead of an path feasibility problem.
Once a basic block, identified by both its block address and VPC, has been discovered and emulated, re-executing that basic block along different paths does not reveal new control-flow structures.
Accordingly, \system symbolically emulates each block at most once.
If additional incoming edges to an already-discovered block are later encountered, these edges are recorded in the CFG, but the block itself is not re-emulated.
By collapsing all incoming paths at block boundaries, \system bounds the number of symbolic states and avoids the path explosion inherent to traditional symbolic execution.

\smallskip
\noindent\textbf{Eliminating reliance on constraint solving.}
In non-obfuscated binaries, CFG recovery can often be performed syntactically by inspecting branch instructions and extracting explicit targets.
Virtualization-based obfuscation breaks this assumption by replacing direct and conditional jumps with indirect jumps whose targets are computed through complex expressions, often involving MBA transformations and values defined far earlier in the program.
Some form of symbolic reasoning is necessary to propagate values to jump sites.
However, traditional DSE accumulates path constraints and queries an SMT solver to determine whether a branch is feasible.
This reliance on satisfiability checking introduces significant overhead and hinders scalability, especially in obfuscated code.

\system eliminates this dependency by observing that CFG recovery does not require path feasibility.
Our goal is not to determine whether a jump can occur, but to enumerate the set of control-flow targets it may resolve to.
Accordingly, when resolving indirect jumps or opaque predicates, \system queries an SMT solver on the recovered symbolic expression \emph{without any accumulated path constraints}.
The solver is used purely as an expression simplifier and value enumerator, not as an input feasibility checker.
This constraint-free usage avoids solver blowups while still recovering all CFG edges.

\begin{figure*}[t]
    \centering
    \includegraphics[width=\textwidth]{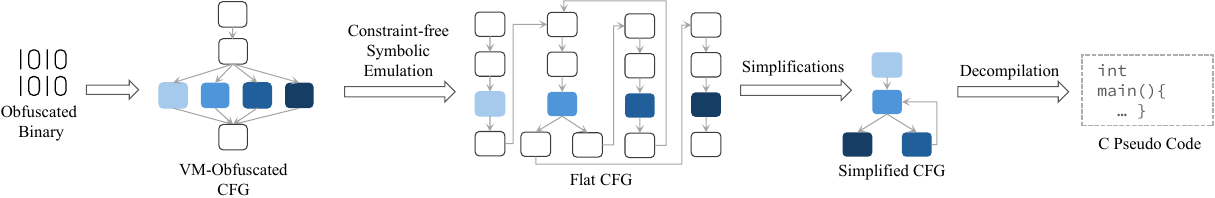}
    \caption{Overview of the analysis pipeline of \system.
%     \jie{It is "Semantics-Preserving Simplification" instead of "Semantic Preserving Simplification"; "VPC-Sensitive CFG Recovery" instead of "CFG Discovery".}
}
    \label{fig:overview}
\end{figure*}

\smallskip
% Together, these insights allow \system to recover complete CFGs from virtualization-obfuscated binaries without relying on execution traces or satisfiable inputs.
% By bounding symbolic state growth and resolving control-flow transfers without path constraints, \system avoids both state explosion and solver complexity that limit traditional symbolic execution.
%
The primary contribution of \system lies in decoupling CFG recovery from path feasibility.
As Figure~\ref{fig:hash_pushan_graph} shows, \system recovers both feasible and infeasible paths of the \texttt{hash} example, yielding a complete CFG (and subsequently, decompiled and simplified pseudocode) suitable for downstream analyses.
% This recovered structure enables effective simplification and decompilation, producing readable C-like pseudocode as shown in Figure~\ref{lst:pushan_decompiled_hash}. \jie{todo}

% To facilitate efficient and effective deobfuscation of virtualization-obfuscated code, we now present \system. We first establishes our threat model, then provide a brief overview and detail individual component of our design.

\subsection{Overview}

\system utilizes a three-stage pipeline (Figure~\ref{fig:overview}) to recover the original control flow and produce decompiled code. % for further analysis.

% \smallskip
% \noindent
% \textbf{Stage 1. VPC-sensitive CFG recovery.}
% \system recovers the control-flow graph (CFG) by emulating the obfuscated code with an abstract state over registers and memory. Each block is uniquely identified by its address \emph{and} virtual program counter (VPC), enabling context-sensitive CFG reconstruction.
% To recover a CFG that closely resembles the original program, \system propagates values through VM instructions to simplify obfuscated jump-target expressions, enabling enumeration of successor addresses for indirect jumps and pruning branches guarded by opaque predicates that evaluate to constants.
% It further employs symbolization to recover additional edges that may be missed during emulation.
% Further details are in Section~\ref{sec:cfg_recovery}.

\smallskip
\noindent
\textbf{Stage 1. VPC-sensitive CFG recovery (Section~\ref{sec:cfg_recovery}).}
\system reconstructs the CFG by emulating the obfuscated binary using an abstract state over registers and memory.
By uniquely identifying blocks by both their native address and virtual program counter (VPC), \system achieves context-sensitive reconstruction.
It propagates values through VM instructions to simplify jump-target expressions, enabling the resolution of indirect jumps and the pruning of constant opaque-predicate branches.
It then uses symbolization to recover edges missed during emulation.
The output of this stage is a CFG with both VM interpreter logic and the original, unobfuscated program's logic.
We call this CFG a \emph{flat CFG}.

\smallskip
\noindent
\textbf{Stage 2. Semantics-preserving simplifications (Section~\ref{sec:simplifications}).}
% The recovered CFG contains redundant low-level logic inherent to the VM architecture.
A flat CFG contains redundant low-level logic inherent to the VM logic.
In this stage, \system applies a series of semantics-preserving transformations on the flat CFG to remove unnecessary artifacts, simplify control flow, and reveal higher-level operations.
\system iteratively applies these simplifications until convergence.

\smallskip
\noindent
\textbf{Stage 3. Decompilation (Section~\ref{sec:decompilation}).}
Finally, the simplified CFG is processed by an enhanced decompiler to generate C-like pseudocode that preserves the recovered logic. Our enhancements include rewriting node identifiers into a decompiler-compatible format, improving function boundary detection despite obfuscated calls, and extending stack pointer tracking to handle non-standard arithmetic.
% These challenges and our corresponding solutions are discussed in detail in .

\subsection{Threat Model}

% To enable efficient deobfuscation of virtualization-obfuscated code, we present \system.
We assume a strong adversary (the virtualization-based obfuscator) with full knowledge of \system's deobfuscation logic and simplification heuristics.
Furthermore, we assume the obfuscator uses a dedicated VPC for each VM instance injected into the binary, as is a standard practice in commercial-grade protectors such as VMProtect and Themida.

Automated binary unpacking is out-of-scope;
\system assumes that the obfuscated binary is either not packed or already unpacked.
While \system must resolve certain opaque predicates (e.g., MBA expressions) to reconstruct the control flow, the exhaustive simplification of these underlying obfuscation primitives is orthogonal to our work.
We treat such techniques as out-of-scope, because specialized solutions like MBA-Blast~\cite{liu2021mbablast} can be used to further simplify \system's output by simplifying or removing MBA expressions.

\section{VPC-Sensitive CFG Recovery}%
\label{sec:cfg_recovery}

Given an obfuscated binary, \system begins by performing VPC-sensitive CFG recovery.
While general-purpose tools for automated VPC identification are generally unavailable, the underlying principles are studied by prior work and not our contribution~\cite{sharif2009automatic,zhang2025vmdoctor}.
\system implements a set of targeted VPC identification heuristics inspired by existing work.
We first describe these VPC identification methods for completeness, then detail our methodology for achieving a complete, VPC-sensitive flat CFG reconstruction.

\subsection{Heuristic VPC Identification}

In virtualization-obfuscated binaries, the VPC typically resides in a register or memory location and serves to index the VM bytecode stream.
The VPC's location varies by obfuscator.
It may be volatile, updated via multi-instruction arithmetic, and transferred between registers (as seen in VMProtect), or stay in a stable location throughout execution (Themida and Tigress).

To handle volatile instances, \system re-evaluates candidate VPC locations at the start of each basic block during symbolic emulation.
Following the methodology in VM-Doctor~\cite{zhang2025vmdoctor}, we identify the VPC by monitoring registers pointing to memory regions with high entropy (a characteristic of VM bytecode regions) and sequential evolution, where the VPC value increases or decreases within some region.
For virtualizers with stable location VPCs, we monitor unique behaviors (e.g., the first memory load from a non-conventional section of a binary, which apply for both Themida and Tigress) and use them to infer VPC locations.
% We first manually identify the storage location in an initial sample to derive a heuristic, such as the first load from a specific binary section, which is then applied to automatically detect the VPC in all subsequent binaries for that obfuscator.
By combining these strategies, \system could successfully identify the VPCs for all evaluated samples across VMProtect, Themida, and Tigress.

While these heuristics are tailored to commercial-strength virtualization-based obfuscators, they are grounded in fundamental properties of virtualization-based obfuscation:
Bytecode indexing and predictable execution flows.
We expect the VPCs of other virtualizers to remain similarly identifiable with minimal manual effort. % through similar structural or behavioral characteristics.

\subsection{Constraint-Free Symbolic Emulation}
\label{sec:sym_emu}

During flat CFG recovery, \system emulates each instruction from the beginning of the obfuscated code and updates an abstract state.
This abstract state includes storage for registers and memory.
\system uses a custom value domain that includes concrete values (e.g., bitvectors and floating points), symbolic values, and a special abstract value \texttt{TOP}, which is created as the result of the union of two or more values when paths merge.

At the beginning of the emulation, we treat all uninitialized values, e.g., environment variables or user inputs, as symbolic.
When encountering external function calls, including standard C library functions and OS APIs, we avoid emulating them.
Instead, we simulate the function's behavior using a function summary when available or return a symbolic value that adheres to the expected calling convention.
% Because common obfuscators typically protect only a single function and do not obfuscate functions called within the to-be-protected function,
We also hook any called unobfuscated functions in the binary to return symbolic values.
This prevents unnecessary analysis and ensures symbolic return values flow correctly into the obfuscated logic.
Note that existing techniques like VMHunt~\cite{xu2018vmhunt} can be applied to automatically identify the boundaries of the virtualized region and automate the hooking process if desired.

\begin{figure*}[htbp]
    \centering
    \begin{subfigure}[b]{0.33\linewidth}
      \centering
      \includegraphics[width=0.60\textwidth]{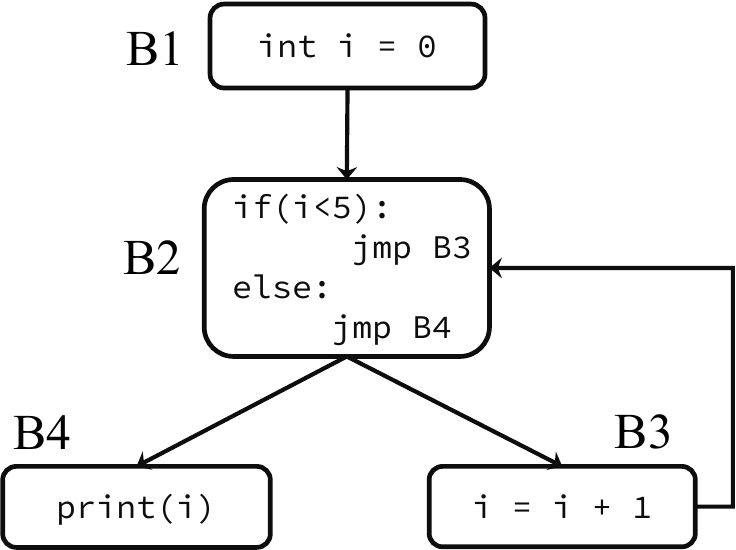}
    %   \caption{Original CFG of an example}
    \caption{}
      \label{fig:sample_original_cfg}
    \end{subfigure}\hfill
    \begin{subfigure}[b]{0.33\linewidth}
      \centering
      \includegraphics[width=0.75\textwidth]{figures/vpc_interpreter.pdf}
    %   \caption{VM-obfuscated CFG of this example. Each handler implements the semantics of one original basic block (B1 to B4).}
    \caption{}
      \label{fig:sample_virtualized_cfg}
    \end{subfigure}\hfill
    \begin{subfigure}[b]{0.33\linewidth}
      \centering
      \includegraphics[width=\textwidth]{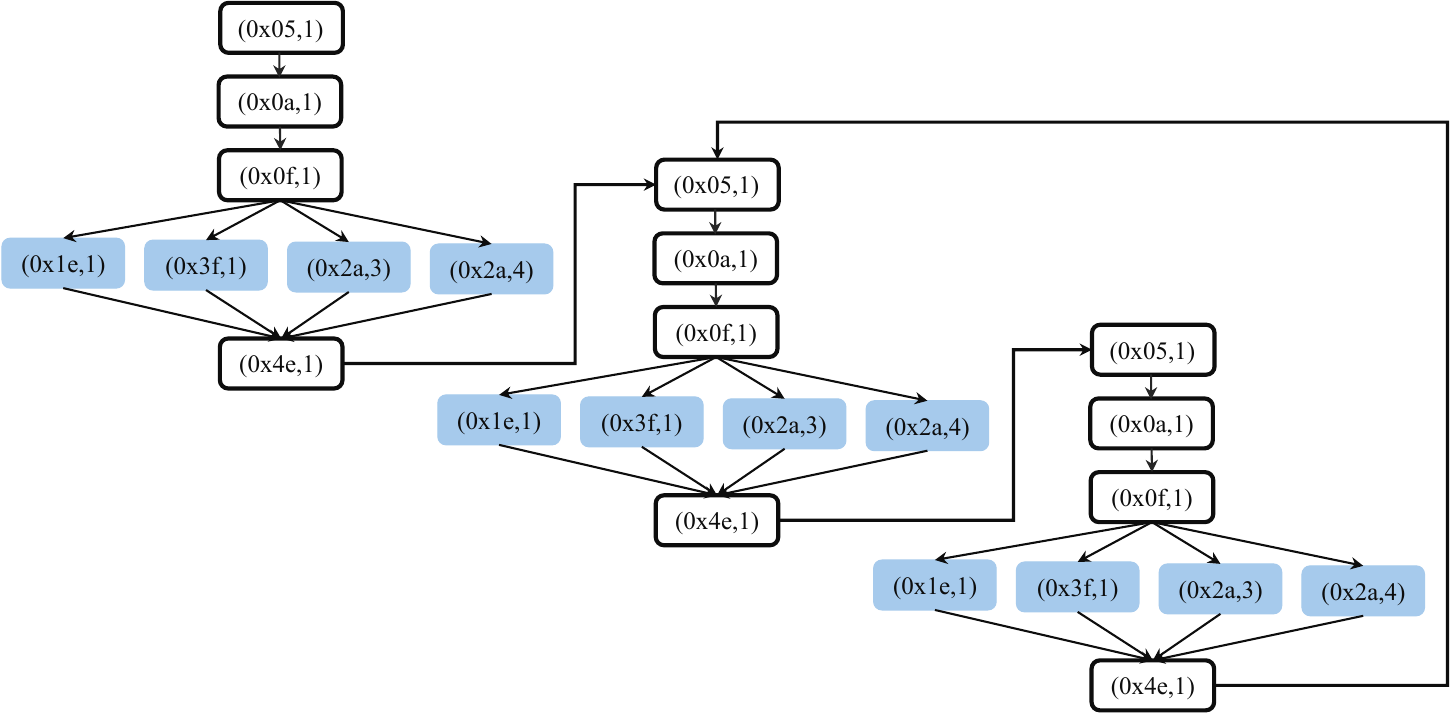}
    %   \caption{\system{}'s initial VPC-sensitive CFG recovery, where a block is identified by (address, VPC).}
    \caption{}
      \label{fig:vpc_sensitive_recovery_stage_1}
    \end{subfigure}
    %\caption{Comparison of original, virtualized, and recovered VPC-sensitive CFGs.}
    \label{fig:first_three_cfgs}
%\end{figure*}

%\begin{figure*}[htbp]
  \centering
  \begin{subfigure}[b]{0.33\linewidth}
    \centering
    \includegraphics[width=0.70\textwidth]{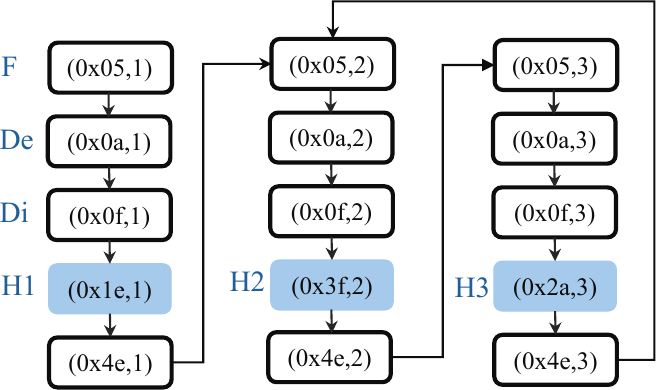}
    % \caption{CFG after eliminating the branches never taken}
    \caption{}
    \label{fig:vpc_sensitive_recovery_stage_2}
  \end{subfigure}\hfill
  \begin{subfigure}[b]{0.33\linewidth}
    \centering
    \includegraphics[width=\textwidth]{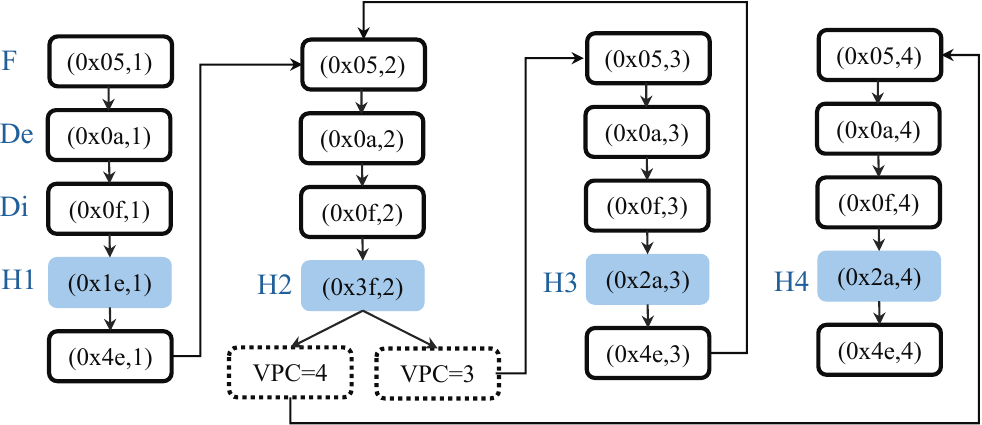}
    % \caption{Symbolization of variable i allows us to recover the exit of the loop}
    \caption{}
    \label{fig:vpc_sensitive_recovery_stage_3}
  \end{subfigure}\hfill
  \begin{subfigure}[b]{0.33\linewidth}
    \centering
    \includegraphics[width=0.60\textwidth]{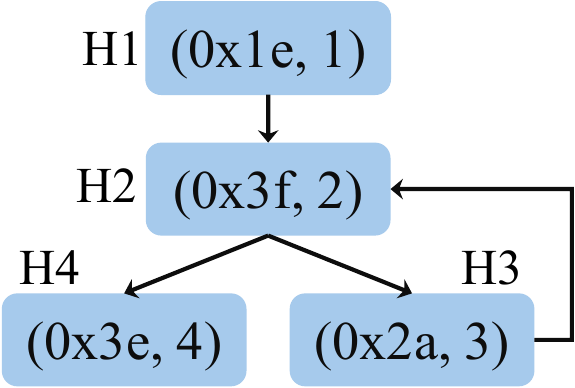}
    % \caption{Final CFG after simplifications, including constant propagation and dead assignment elimination, which remove the VM interpreter's remnants.
    %First constant propagation is performed, which propagates the byte codes through the program. Once this happens most of the instructions in the fetch, decode execute blocks become redundant and can be eliminated with the semantics preserving transformations such as dead assignment elimination
    % } %
    \caption{}
    \label{fig:vpc_sensitive_recovery_stage_4}
  \end{subfigure}
%   \caption{Exemplary code (a) that has been obfuscated (b) and overview of how \system{} deobfuscates it step-by-step (c-f).}
    \caption{The \system deobfuscation workflow applied to a virtualized example. (a) The original CFG of the target program. (b) The VM-obfuscated CFG, where individual handlers encapsulate the semantics of original basic blocks ($B_1$–$B_4$). (c) Initial VPC-sensitive CFG recovery. (d) CFG refinement after eliminating infeasible branches and unreachable paths. (e) Recovery of the loop exit condition following the symbolization of variable $i$. (f) Final CFG after simplifications.}
  \label{fig:second_three_cfgs}
\end{figure*}

During emulation, \system creates a new node for each newly discovered basic block and adds a new edge between the source node and the new node in the CFG when a control flow transfer happens.
To implement VPC sensitivity, we uniquely label each node using a tuple of (block address, VPC).
We refer to this tuple as the ID of this node.
After the initial VPC-sensitive CFG exploration, our graph looks like Figure~\ref{fig:vpc_sensitive_recovery_stage_1}.

During each exploration round, \system visits each VPC-sensitive block at most once; exploration terminates when no new (address, VPC) pairs are discovered.
After each exploration round, symbolization (Section~\ref{sec:symbolization}) may introduce additional symbolic values, which enables the discovery of previously missed control-flow edges in subsequent rounds.
This iterative process gradually expands the recovered CFG.
% This means that when there's a loop in the obfuscated program, we only follow it once.

\smallskip
\noindent
\textbf{Pruning branches for direct conditional jumps.}
Virtualization-based obfuscators frequently apply opaque predicates to thwart static analysis effort, and a common use case for opaque predicates is to convert an unconditional jump into a conditional jump where one of its branches is always satisfiable~\cite{kochberger2021sok}.
The end result is an excessive number of redundant control-flow transfers in the obfuscated code.
\system simplifies the CFG and prunes the always-unsatisfiable branches using the following rules:

\begin{enumerate}[itemsep=0pt, parsep=1pt, topsep=2pt, partopsep=2pt]
    \item
        Pruning branches with always-false conditions (opaque predicates).
        \system evaluates each branch condition in isolation, without any path constraints, using an SMT solver as an expression simplifier.
        This is because static simplification rules cannot handle complex opaque predicates.
        We note that existing MBA simplification methods can be applied~\cite{liu2021mbablast,reichenwallner2022simba,reichenwallner2023gamba}.

    \item
        Keeping all branches with symbolic or user input-dependent conditions.
\end{enumerate}

\noindent
After pruning, our graph looks like Figure~\ref{fig:vpc_sensitive_recovery_stage_2}, where all the flattened handler copies unrelated to the current VPC are eliminated along with fake branches that depend on opaque predicates (not shown in the graph).

\smallskip
\noindent
\textbf{Resolving obfuscated indirect jump targets.}
In traditional binary disassembly, determining the target of a conditional jump is straightforward, as the target address is embedded directly in the instruction itself (e.g., \code{jz \$+5}).
In contrast, obfuscated binaries often replace direct jumps with indirect jumps, where the target address is computed.
We use symbolic emulation to recover the intended jump targets by executing the relevant instructions and tracking how the jump target is calculated.
Obfuscators often apply MBA transformations to make these expressions harder to interpret. % by hiding both the control conditions and the resulting addresses.
Listing~\ref{fig:mba_example} shows an example.
To resolve such indirect jumps, \system applies the same strategy used for pruning opaque branches:
It evaluates the recovered expression in isolation, without applying any path constraints.
We stress that the goal is not to determine which input would trigger the jump, but to enumerate all possible outputs of the expression---that is, the set of target addresses it may evaluate to.
This mirrors traditional disassembling techniques, which recovers jump destinations rather than reasons about the conditions under which they are taken.

\begin{listing}[b]
    \centering
    \begin{minipage}{\linewidth}
    \begin{lstlisting}[
        basicstyle=\fontfamily{pcr}\scriptsize,
        xleftmargin=10pt,
        xrightmargin=5pt,
    ]
~(~(0xfffffffff + (0x0 .. ~(if ~(0x5 + ~input) == 0x0 then 1 else 0))) | -5369364481)
+ ~((0xfffffffff + (0x0 .. (if ~(0x5 + ~input) == 0x0 then 1 else 0))) | -5369367216)
    \end{lstlisting}
    \end{minipage}
% \caption{An example obfuscated expression that calculates the jump target based on the value of a symbolic variable \texttt{input}.}%
\caption{Example of an obfuscated jump target calculation using symbolic variable \texttt{input}.}
\label{fig:mba_example}
\end{listing}

\subsection{Symbolization}
\label{sec:symbolization}

% Handling missing branches - Symbolizer
Pruning may lead to missing branches during the first analysis pass of a basic block.
Consider the exit branch of a loop that iterates for a constant number of iterations.
The obfuscated guard condition of the exit branch will only evaluate to one value (going to the beginning of the loop) unless the execution reaches the last iteration.
\system would incorrectly prune the exit branch, leading to an incomplete flat CFG.
To address this problem, \system introduces \emph{symbolization} to discover and symbolize non-constant variables.

\begin{enumerate}[itemsep=0pt, parsep=1pt, topsep=2pt, partopsep=2pt]
    \item
        When emulating each instruction, \system keeps the values in registers and memory in an abstract state.
        \system also resolves the addresses of load and store instructions. % so that we can perform the emulation accurately.

    \item
      When two paths merge during emulation (e.g., a back edge reaching the loop header), \system merges the abstract states for the two paths, where constant but distinct values at the same location are merged into TOP (an unconstrained symbolic value).
        % We only do this for constants, as variables are already symbolized. %we do not want to over-symbolize by replacing already symbolized variables.
        For dealing with opaque predicates, we use an SMT solver (without constraints) to check if an expression evaluates to a constant.
        We also save the information about which locations have been symbolized and at which block ID.

    % \item
    %     When symbolizing a location, we must ensure that it is not already symbolic, such as from a previous symbolization or an uninitialized variable.

    \item
        Symbolization terminates once the set of symbolic locations reaches a fixed point (i.e., no new symbolic locations are introduced through state merging).

    \item
      When performing CFG recovery (Section~\ref{sec:sym_emu}) again, \system can use the saved information regarding symbolized locations.
        % If \system encounters a previously saved block ID during the emulation, we symbolize the values in the abstract state for that block by replacing them with a TOP variable and continue emulating with this new state with symbolized values.
\end{enumerate}

\noindent
Symbolization allows for symbolic values to flow into the obfuscated conditions of conditional jumps and discover both branches.
% This is also helpful when jump targets can no longer be inferred from the instruction itself, as is the case for VMProtect-obfuscated binaries, which converts conditional jumps into indirect jumps.
% In binaries protected using VMProtect, where conditional jumps are converted into indirect jumps, the jump targets can no longer be inferred from the instruction opcode itself.
% If this were a constant conditional check like that of a loop running a constant number of times, we would miss the exit branch.
Symbolization allows \system to retrieve all possible jump targets for these types of indirect jumps by symbolizing the constant value that the condition checks for and exploring both branches.

Figure~\ref{fig:vpc_sensitive_recovery_stage_2} examplifies symbolization, where initializing \code{i=0} makes the loop-exit guard false and prunes the edge.
After symbolization, \code{i} becomes symbolic, and both successors are explored (Figure~\ref{fig:vpc_sensitive_recovery_stage_3}).

\smallskip
\noindent
\textbf{Increasing scalability by limiting symbolic solves.}
Many obfuscated expressions can be simplified to constants using an SMT solver, but it is time-consuming and does not scale.
We choose a granularity where we resolve the constants by deciding to not solve for all possible expressions, and only solving for certain types of expressions, e.g., the addresses for loads and stores, because these expressions are usually for loading VM bytecode from the bytecode region.
This way, we reduce the time spent in solving and still get the constants that are essential for simplifying away the VM machinery.

\smallskip
\noindent
\textbf{Mitigating the NP-hardness of path satisfiability.}
Existing approaches~\cite{coogan2011deobfuscation,yadegari2015generic,salwan2018symbolic,xu2018vmhunt} first collect an execution trace, then accumulate path constraints, and finally query the solver for inputs that satisfy yet-unexplored paths.
In contrast, \system sidesteps full path satisfiability by adopting three lightweight techniques:
\begin{enumerate}[label=(\roman*)]
    \item \textbf{Constraint-free simplification.}
          To enumerate jump targets, we pass the SMT solver only the target expression, omitting all accumulated path constraints.
          This keeps each query lightweight and tractable.

    \item \textbf{Symbolization.}
          Whenever two paths merge, any differing concrete value is replaced with a fresh symbolic variable (e.g., the initial value~0 of \texttt{hash} in Listing~\ref{lst:pre_image_hash}), dramatically shrinking the expressions handed to the solver.

    \item \textbf{Single-iteration loop handling.}
          Each loop is executed at most once during symbolic emulation, preventing complex expressions from forming—which would otherwise be harder for the solver to handle.
          For instance, in Listing~\ref{lst:pre_image_hash} the symbolized \texttt{hash} variable is computed over only one iteration; although this yields an approximated expression, it is sufficient for recovering all jump targets, as we do not care about the actual feasibility.

\end{enumerate}
These three tactics suffice in practice, but they can be combined with modern MBA simplifiers~\cite{liu2021mbablast,reichenwallner2022simba,reichenwallner2023gamba} to reduce complex symbolic expressions into forms from which the jump targets can directly be inferred.

\section{Semantic-preserving Simplifications}%
\label{sec:simplifications}

A flat CFG is complete but insufficient for decompilation (or many other downstream analyses), because it contains many redundant instructions that correspond to VM interpreters.
\system iteratively applies a series of semantic-preserving simplifications on the flat CFG until reaching a fixpoint and producing a simplified CFG that resembles the one in the original unprotected program.

\subsection{List of Simplifications}%
\label{sec:list_of_simplifications}

\system uses the following simplifications. % An example can be found in Figure~\ref{fig:simplfication}
% More transformations can be added depending on the specific obfuscations used in the binary.

\smallskip
\noindent
\textbf{S1. Standard simplifications.}
\system applys constant propagation across the flat CFG.
\system treats VM bytecode regions as constant (read-only), allowing bytecode values to propagate.
This propagation renders much of the VM interpreter machinery redundant (e.g., handler dispatch logic that loads the handler address from the bytecode region) and eliminatable by subsequent simplifications.

\system then applies dead assignment elimination to remove redundant bytecode-processing logic.
This step also eliminates opaque predicates by resolving their guard conditions.
% \system performs this analysis at the block level (registers and stack variables) and the function level (registers only).
% This also helps remove instructions related to opaque predicates, where we identified the final value of the opaque predicate guard condition after constant propagation.
% This analysis is performed at the block level (registers and stack variables) and the function level (registers only).

Lastly, \system performs basic arithmetic simplifications, e.g., grouping together the arithmetic operations performed on the same register.
This simplification pass reduces the arithmetic operations left after dead assignment elimination.
Figure~\ref{fig:semantic_preserving_code_simplifications} in the appendix shows an example.

\smallskip
\noindent
\textbf{S2. Redundant stack variable removal.}
A common pattern in simplified code is a value being stored at a stack location and immediately loaded into a register (and the stack location is never loaded after).
\system simplifies this pattern by removing the redundant stack write.
% However, before replacing the load, as in Figure~\ref{fig:semantic_preserving_code_simplifications} in the appendix, we must ensure that no other value is stored at the same location before the final load.
To avoid issues with pointer aliasing, we only apply this simplification within a basic block.

\smallskip
\noindent
\textbf{S3. Removing self-defining variables in loops.}
\system further removes self-referential stack and global variables that persist after dead assignment elimination.
These variables, often remnants of loop counters, are pruned if their only use is self-incrementing or self-referencing.
To maintain semantic integrity, \system preserves any variable that influences control-flow guards or serves as a function argument.

% %
% Some unused or self-referential stack and global variables are not eliminated even after dead assignment elimination.
% These are typically present in loops, acting as counters, where their only use is at the beginning of the loop.
% We remove such variables by checking if the value is only used by itself, i.e., loop incrementing; if it ends up in condition checks or function arguments, then we do not remove it.

\smallskip
\noindent
\textbf{S4. Obfuscator-specific simplifications.}
Both VMProtect and Themida use transformations that are not simplified by previous techniques.
For example, Themida converts every conditional jump into two conditional jumps.
The first jump computes the check and sets a global variable.
The second conditional jump checks the value of the global variable and chooses the correct branch.
\system includes obfuscator-specific simplifications to simplify such patterns.
% Similarly, we implemented peephole optimizations for structural and instruction-level simplifications.

\subsection{Differences to Prior Work}%
\label{sec:simplification_differences}

While prior work~\cite{yadegari2015generic} has implemented some simplifications (S1), we note some key differences below.

\begin{itemize}[nosep]
    \item
    Unlike prior work that operates on individual traces~\cite{yadegari2015generic}, \system analyzes a complete CFG.
    This global perspective mitigates the risk of over-simplification inherent in trace-based approaches, which rely on precise, end-to-end taint tracking of user input to maintain correctness.
    \system remains input-agnostic and does not require a-priori knowledge of input sources or their propagation.
    % \system works on a complete CFG while prior work~\cite{yadegari2015generic} only simplifies traces.
    % Consequently, as discussed in the paper, the solution suffers from over-simplification and requires accurate taint tracking of user input across the entire trace to avoid over simplification.
    % \system does not require knowing where user input is or how it is used.

    % \item
    % Some simplifications in the original solution are rule-based and may not be sound.
    % \system only performs sound and semantic-preserving simplifications.

    \item
    Prior work may yield different simplification results when simplifying multiple traces, which makes merging traces into a CFG difficult.
    \system simplifies the CFG directly, so it does not need to reconstruct it.

    \item
    \system implements more simplifications (S2, S3), as well as obfuscator-specific simplifications (S4).

\end{itemize}

\section{Decompilation}%
\label{sec:decompilation}

\system takes the simplified CFG from the previous stage and decompiles the entire CFG using a customized open-source binary decompiler.
We discuss key steps in this section.

% \medskip
% \noindent
% \textbf{Rewriting node addresses.}
% Before decompilation, \system rewrites all node IDs from three-tuples to integers so that the decompiler can accept the CFG as input.

\smallskip
\noindent
\textbf{Identifying function boundaries.}
Most binary decompilers only decompile a function at a time.
However, the function boundaries are unclear in obfuscated binaries, especially the function return sites, as there are many fake call and return statements that do not jump to or return from any functions. 
Additionally, some obfuscators will rewrite \texttt{call} instructions into \texttt{jmp} instructions to further obfuscate the control flow. 
While most bogus call and return instructions are removed during the simplification stage, there still remain some jump or return statements that actually transfer control flows. 
To address this, \system rewrites return statements and jump statements that go to functions inside the unobfuscated code of the binary into call statements.

\smallskip
\noindent
\textbf{Enhancing stack pointer tracking.}
Decompilers usually assume that specific registers (e.g., \texttt{rsp} on X86-64) are always stack pointers.
This assumption does not always hold with obfuscated binaries, where \texttt{rsp} can be repurposed to hold generic values, causing the decompiler to fail to track or recover stack variables.
% For example, in VMProtect, the virtual stack pointer is never rsp, so if we fail to track this virtual stack pointer register, we will incorrectly track the stack pointer. 
Obfuscators may also insert code to conduct unconventional arithmetic operations, such as Negation, Xor, or MBA expressions, on stack pointer registers. This also curtails the stack pointer tracking that decompilers perform.
% Typically, you would not expect operations such as not, xor, etc., to be performed using the stack pointer. 
Figure~\ref{fig:sp_altering} shows such an example.
We enhanced the open-source decompiler to discover the actual stack pointer registers and added support in stack pointer tracking for unconventional arithmetic operations and simplification rules. 

\begin{figure}
    \center
    \begin{minipage}{\linewidth}
    \begin{lstlisting}[basicstyle=\ttfamily\scriptsize, aboveskip=0pt, belowskip=0pt,xleftmargin=15pt,xrightmargin=5pt]
not(not(rsp)+0x18) ==> rsp-0x18
    \end{lstlisting}
    \end{minipage}
\caption{An example of using an unconventional operation, Negation, when computing stack pointer offsets.}%
\label{fig:sp_altering}
\end{figure}

%-------------------------------------------------------------------------------
\section{Evaluation}%
\label{s:eval}
%-------------------------------------------------------------------------------

We first evaluate the effectiveness and correctness of \system on a diverse set of binaries for which we have the ground-truth (original source code or CFG).
Our experiments include analyzing the deobfuscation and decompilation output (Section~\ref{sec:eval_effectiveness}), comparing the deobfuscated CFGs to the original CFGs and quantifying their similarity (Section~\ref{sec:eval_cfgsimilarity}), and validating the semantic correctness of the deobfuscated code (Section~\ref{sec:eval_semcorrectness}).
We then show with a case study that \system can achieve logic recovery using deobfuscated and decompiled code from a malware sample (Section~\ref{sec:eval_codereuse}).
Lastly, we evaluate the effectiveness of \system on custom virtualization-based obfuscation implementations in binaries from Capture-the-Flag (CTF) competitions, for which we do not have ground truth (Section~\ref{sec:eval_customvm}).

\subsection{Dataset}

To thoroughly evaluate \system, we constructed a dataset comprising multiple program categories.
It includes six Windows malware samples used in prior work~\cite{yadegari2015generic}, four open-source projects, four hand-crafted binaries, five challenges from Capture the Flag (CTF) competitions with custom VM implementations, 1,000 Tigress-generated virtualized binaries, and a real-world VMProtect-obfuscated executable.

\smallskip
\noindent
\textbf{Real malware.}
Our dataset includes six Windows malware samples (source code available) used in prior work~\cite{yadegari2015generic}.
The malware contain typical malicious behaviors, e.g., backdoors and spreaders, demonstrating \system's intended use case of malware analysis.
The original versions of the samples totaled 1,125 instructions (see Table~\ref{tab:merged-eval-alternative}), which expanded to 531{,}790 instructions under VMProtect and over 5 million under Themida.
For comparison, the 1993 remake of \textit{Doom} contains under 4 million instructions.

\smallskip
\noindent
\textbf{Open-source software.}
We selected one function each from four widely used libraries---zlib (\texttt{gz\_read}), curl (\texttt{glob\_url}), libpng (\texttt{png\_decompress\_chunk}), and SDL (\texttt{ClosePhysicalCamera}).
These cover reading bytes from a file, URL parsing, decompressing data, and device event handling, respectively.

\smallskip
\noindent
\textbf{Synthetic samples.}
We created four synthetic test binaries to evaluate obfuscation scenarios not covered by the malware samples.
These include: 1) \textit{if-cond}, which contains a simple input-dependent conditional jump, 2) \textit{hash}, which implements the motivating example in Listing~\ref{lst:pre_image_hash}, 3) \textit{const-loop}, which features complex loops that challenge DSE techniques, and 4) \textit{huffman} for comparability to prior work.

\smallskip
\noindent
\textbf{Custom VMs from CTFs.}
We included five CTF challenges with bespoke virtualization-based obfuscation to test \system's generalizability.
These challenges all implement \emph{custom} VM architectures, and \system was able to successfully deobfuscate all of them.
In one of these challenges, the flag value was visible directly in the decompiled result, demonstrating \system's ability to recover high-level semantics even from heavily obfuscated, custom virtual machines.

\begin{figure*}[ht]
    \centering
    \begin{adjustbox}{max width=0.8\textwidth}
    \begin{subfigure}[b]{0.32\linewidth}
        \centering
        \includegraphics[width=\linewidth]{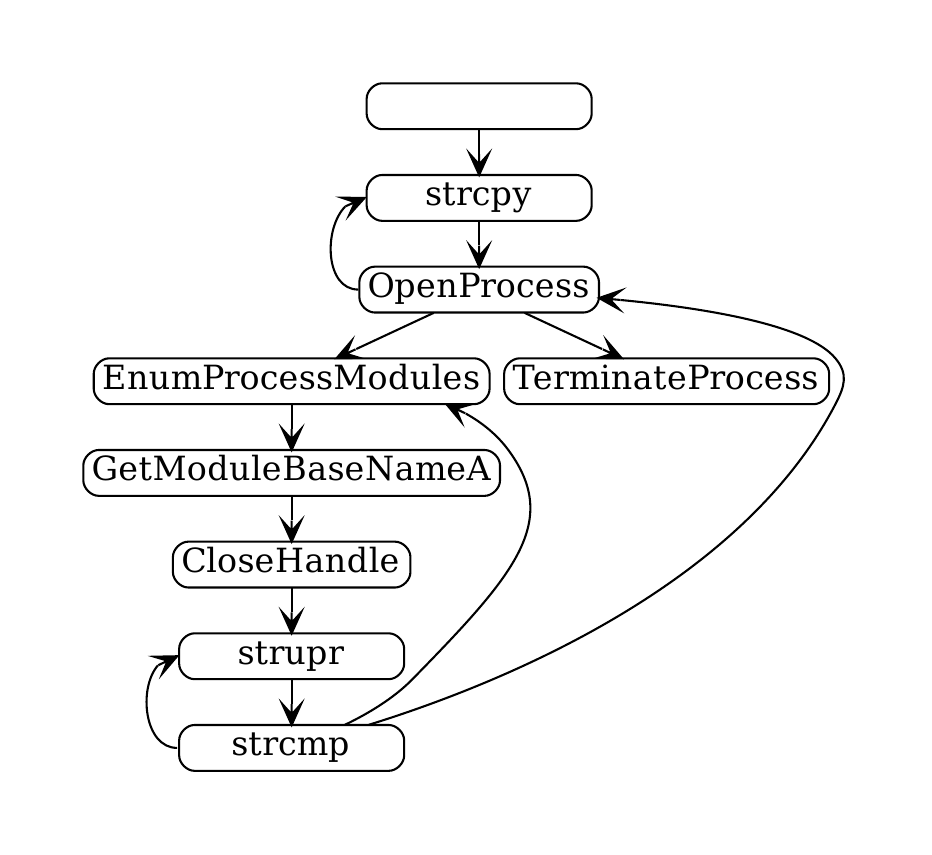}
        \caption{Prior work's deobfuscated call graph of Netsky~\cite{yadegari2015generic}, with missing blocks and API calls.}
    \end{subfigure}
    \hfill
    \begin{subfigure}[b]{0.32\linewidth}
        \centering
        \includegraphics[width=\linewidth]{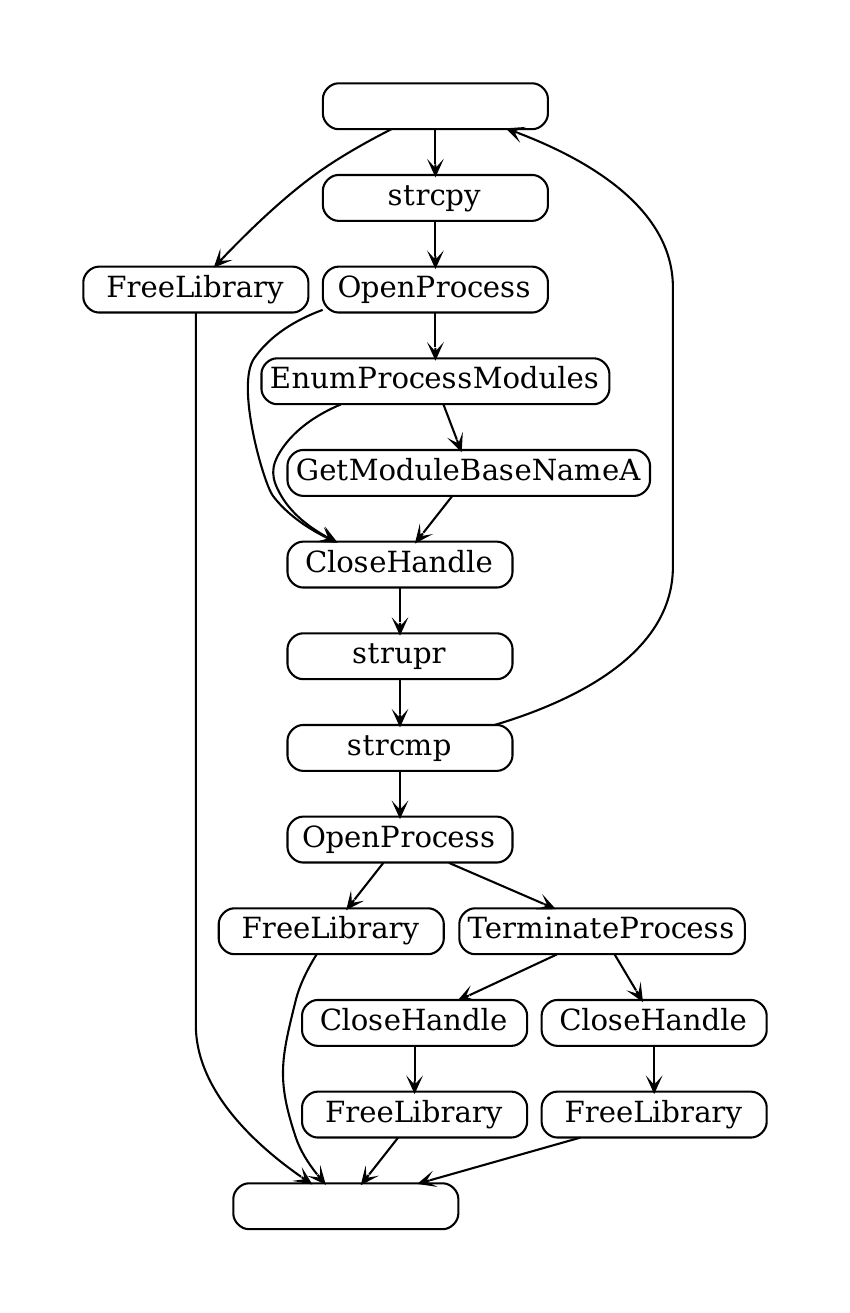}
        \caption{Original call graph of Netsky.}
    \end{subfigure}
    \hfill
    \begin{subfigure}[b]{0.32\linewidth}
        \centering
        \includegraphics[width=\linewidth]{figures/deobf_super_graph.pdf}
        \caption{Deobfuscated call graph by \system.}
    \end{subfigure}
\end{adjustbox}
    \caption{Comparison of Netsky's original function call graph (middle), prior work's deobfuscated function call graph (left), and the deobfuscated, complete call graph by \system (right).}%
    \label{fig:netsky_cfgs}

\end{figure*}

\smallskip
\noindent
\textbf{Large-scale correctness dataset.}
For Tigress, we used the random function generator to produce $1,000$ hash functions, which we then obfuscated using virtualization along with the following configurations:
\begin{itemize}[nosep]
    \item \textbf{VM-1}: Stack-based VM with switch-case dispatch.
    \item \textbf{VM-2}: Opaque predicates, duplicate opcodes, direct dispatch, super operators, and additional obfuscations.
    \item \textbf{VM-3}: Virtualization with bogus functions, implicit flows, and opaque predicates.
\end{itemize}
Tigress operates at the source level, so we compiled the obfuscated source code to generate the final obfuscated binaries.

\smallskip
\noindent
\textbf{A real VM-obfuscated sample.}
Lastly, we evaluated \system on a real-world VMProtect-obfuscated binary.
\system successfully recovered the logic of obfuscated functions, allowing us to determine the true nature of the binary.

\subsection{Experiment Environment and Design}
We ran all experiments on a Kubernetes cluster with 2.30GHz Intel Xeon CPUs, allocating up to 80GB of RAM per sample.
We selected two popular commercial binary obfuscators: VMProtect 3.5.0~\footnote{The authors of this paper cannot purchase newer versions of VMProtect for compliance reasons.} and Themida 3.1.1.0.
Both obfuscators offer virtualization-based obfuscation and other protections; for our evaluation, we disabled all but virtualization-based obfuscation.
All our Themida-obfuscated samples use the Fish VM variant.
To assess generalizability across different VM configurations, we also tested two samples each with the Tiger and Dolphin VM variants.
In all cases, \system achieved the same CFG similarity scores across variants.

Each VMProtect and Themida sample in our dataset was compiled with and without the respective obfuscator SDKs, resulting in three binaries per sample: one original (used as ground truth) and two obfuscated binaries.

We implement a prototype of \system on angr~\cite{shoshitaishvili2016sok}.
Our prototype mainly works with VEX IR during CFG recovery and simplification, then transitions to angr IL (AIL) for decompilation.
Our heuristic-based VPC finder automatically identifies the VPC for VMProtect, Themida, and Tigress.
For CTF challenges, we manually specify the VPC locations.

\smallskip
\noindent
\textbf{Comparison to prior work.}
Yadegari et al.~\cite{yadegari2015generic} did not fully released their tool.
Upon contacting them, we learned that the concolic execution component of their tool is not publicly available.
Instead, they only released the final CFGs for some of the samples.
These CFGs consist solely of the graph structure, without any accompanying deobfuscated code, which is insufficient for accurately comparing similarities.
VMHunt~\cite{xu2018vmhunt} provides source code but lacks runnable samples, test binaries, or documentation needed for reproducing their results.
Salwan’s framework~\cite{salwan2018symbolic} targeting Tigress binaries is available and functional.
We use it as a baseline for evaluating our approach on the Tigress binaries of our dataset.
% The virtual program counter in Themida is placed at a fixed location for each binary, so identifying the location once is enough.
% Existing work for automatically identifying the VPC can also be used in this annotation process~\cite{todo}.

% \usepackage{graphicx}
% Adjust the space above and below the caption
\setlength{\abovecaptionskip}{5pt} % Space above the caption
\setlength{\belowcaptionskip}{5pt} % Space below the caption
\begin{table*}[tb]
    \centering
    \caption{
    	    Evaluation results for open source software, malware and synthetic programs showing the impact of virtualization-based obfuscation.
    VMProtect and Themida expand binaries by several orders of magnitude.
    CFG similarity is measured using our enhanced graph edit distance metric, showing high CFG similarity across all samples.
    }\label{tab:merged-eval-alternative}
    % \resizebox{\textwidth}{!}{%
    {\footnotesize
    \begin{tabular}{ll|rrr|r|rr|rr|rr}
        \hline
        \multirow{2}{*}{Program} & \multirow{2}{*}{Function} &  \multicolumn{3}{c|}{Instructions} &  \multicolumn{1}{c|}{LoC} & \multicolumn{2}{c|}{DLoC} &  \multicolumn{2}{c|}{Similarity on VMP} &  \multicolumn{2}{c}{Similarity on TH} \\ 
        \cline{3-12}
        & & Orig & VMP & TH & Orig & VMP & TH & Pushan & Yadegari & Pushan & Yadegari \\
        \hline
        zlib             & gz\_read              & 188 & 302,697 &   669,613 &  70 & 417 & 162 & 100\% &   -- & 100\% &   -- \\
        curl             & glob\_url             & 133 & 169,925 &   704,548 &  42 & 137 &  73 &  86\% &   -- &  76\% &   -- \\
        libpng           & png\_decompress\_chunk& 274 & 378,108 & 1,027,454 & 134 & 472 & 154 &  69\% &   -- &  47\% &   -- \\
        sdl              & ClosePhysicalCamera   & 139 & 200,077 &   831,575 &  50 & 367 & 305 &  81\% &   -- & 100\% &   -- \\
        \hdashline
        netsky           & main                  & 150 & 162,140 &   679,448 &  33 & 146 & 123 &  96\% & 33\% &  96\% & 19\% \\
        hunatcha         & InfectDrives          &  99 &  50,999 &   791,110 &  22 &  51 &  25 & 100\% & 30\% & 100\% & 23\% \\
        cairuh           & kazaa\_spread         & 186 & 142,148 & 1,466,296 &  28 & 147 & 118 & 100\% &   -- &  93\% &   -- \\
        blaster          & blaster\_spreader     & 140 &  99,389 &   773,785 &  64 & 130 &  87 &  92\% & 49\% & 100\% &   -- \\
        newstar          & InfectExes            &  81 &  33,760 &   489,129 &  18 &  40 &  22 & 100\% & 53\% & 100\% & 18\% \\
        newstar          & Backdoor              & 116 &  43,354 &   848,283 &  31 &  27 &  48 & 100\% &   -- & 100\% &   -- \\
        \hdashline
        huffman          & create\_huffman\_codes& 205 & 106,001 &   851,284 &  30 & 241 & 148 &  82\% & 33\% &  92\% & 31\% \\
        hash             & main                  &  51 &  59,976 &   273,381 &  12 &  75 &  40 & 100\% &   -- & 100\% &   -- \\
        if-cond          & main                  &  15 &  37,951 &   163,950 &   6 &  11 &  13 & 100\% &   -- & 100\% &   -- \\
        const-loop       & main                  &  23 &  43,354 &    61,943 &   5 &  53 &  17 & 100\% &   -- & 100\% &   -- \\
        \hline
    \end{tabular}
    }
    \caption*{\scriptsize
        \textbf{DLoC:} Decompiled Lines of Code (LoC) after Pushan's deobfuscation. 
        \textbf{Similarity:} CFG similarity score computed using enhanced GED. 
        \textbf{Orig:} Instructions / Lines of Code (LoC) in the \emph{original}, unobfuscated binary. 
        \textbf{VMP:} VMProtect. 
        \textbf{TH:} Themida.
    }
\end{table*}

% \system took 72 minutes for the VMProtect-obfuscated binary and 166 minutes for the Themida-obfuscated binary, successfully recovering both paths in the function.

\subsection{Effectiveness}
\label{sec:eval_effectiveness}

To show that \system handles real-world malware and aids reverse engineering, we analyzed multiple aspects of decompilation.

\smallskip
\noindent
\textbf{Comparison against original source code.}
Table~\ref{tab:merged-eval-alternative} compares lines of decompiled code against source code for programs whose original code is available.
Some binaries, such as \textit{huffman}, show significant line expansion due to obfuscated conditional branch representation.
VMProtect, for instance, replaces conditional jumps (\textit{jz}, \textit{jle}) with flag bit calculations.
This spreads condition checks across multiple instructions that our semantics-preserving transformations cannot simplify to the original form.
Resolving this requires mapping instruction sequences back to original conditional jumps.
We leave it as future work as it does not impact our core technique.

For Tigress samples (Table~\ref{tab:tigress-results}), the line count increase stems from limited code-level simplifications rather than control-flow complexity.
\system outputs one arithmetic operation per line for Tigress, while the original source combines multiple operations per line.
This increases line count despite equivalent underlying logic.

\begin{figure*}[tb]
    \centering
    \includegraphics[width=0.7\textwidth]{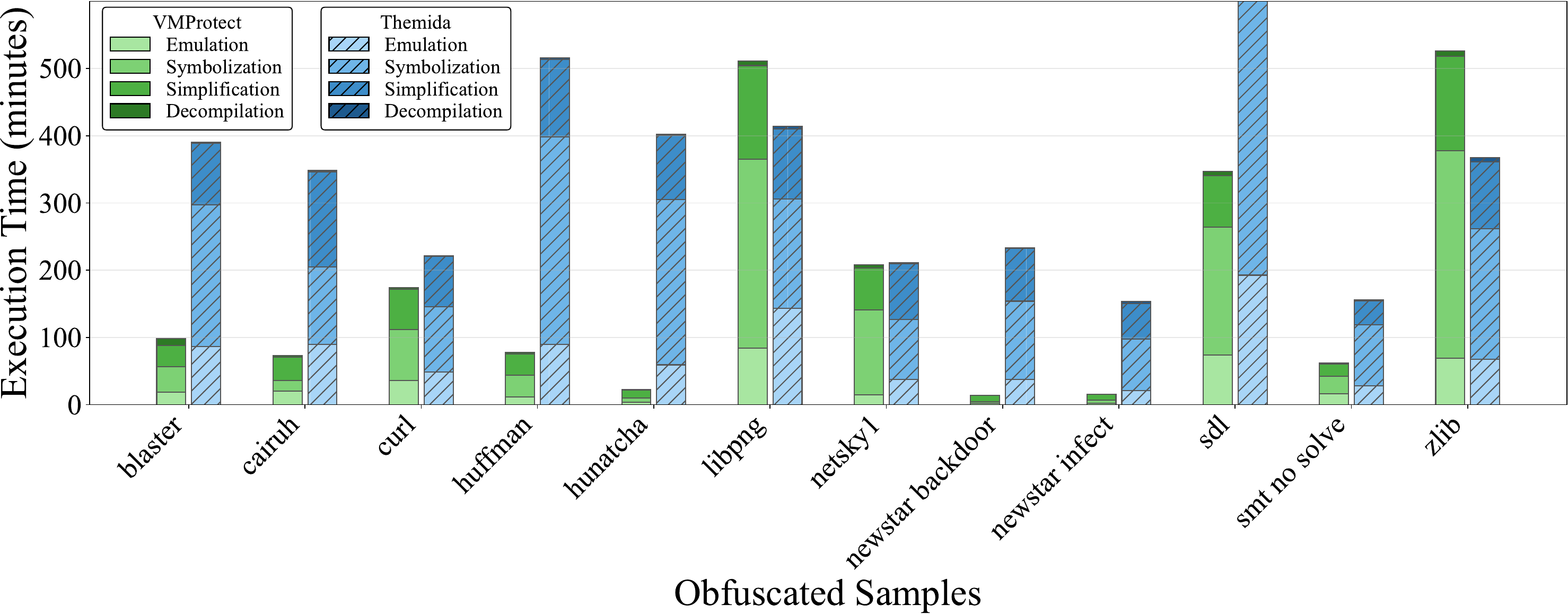}
    \caption{Time taken by different stages of \system. Time for sdl-Themida is excluded from graph due to scale. (Symbolization: 512.72, Simplification: 109.52, Decompilation: 4.59) }
    \label{fig:analysis_stages}
\end{figure*}

\smallskip
\noindent
\textbf{Runtime performance.}
Figure~\ref{fig:analysis_stages} shows the runtime breakdown by stage.
Themida-obfuscated binaries take longer to analyze because they contain more instructions.
\system's performance scales with binary size and the number of constant-guarded nested loops, requiring one symbolization run per such condition.
Our largest sample, \textit{huffman} (Themida), contained 851,284 instructions and took ~10 hours to analyze.
Other large samples—\textit{hash} (Themida, 273,381 instructions) and \textit{huffman} (VMProtect, 106,001 instructions)—completed in under 4 and 2 hours, respectively.

\begin{table}[tb]
    \centering
    \caption{CTF Challenge Analysis. DLoC = lines of decompiled code. ``API'' indicates whether the recovered CFG contained the API trace from execution. ``Writeup'' indicates whether the recovered CFG matched functionality described in publicly available CTF writeups.}
    \label{tab:ctf-api-trace}
    {\footnotesize
    \begin{tabular}{lccc}
    \toprule
    \textbf{Challenge} & \textbf{DLoC} & \textbf{API} & \textbf{Writeup} \\
    \midrule
    reduced-reduced-instruction-set-1 & 158 & \checkmark & \checkmark \\
    Discount VMProtect & 86  & \checkmark & \checkmark \\
    simple-vm & 110 & \checkmark & \checkmark \\
    Highly Optimized & 167  & \checkmark & \checkmark \\
    vmwhere1 & 1768 & \checkmark & \checkmark \\
    \bottomrule
    \end{tabular}
    }
\end{table}

\subsection{Correctness: CFG Similarity}
\label{sec:eval_cfgsimilarity}

We first measure the graph edit distance (GED) of our deobfuscated CFG to the original CFG.
Previous work~\cite{yadegari2015generic} utilized the Hu GED algorithm~\cite{hu2009large} to measure the approximate similarity of their graphs.
We found this GED estimation algorithm inapplicable for two reasons.
First, the Hu algorithm relies on the presence of x86 instructions to improve its results by matching opcodes---\system does not produce x86 instructions.
Second, the Hu algorithm often incorrectly reports identical graphs as different.
For interested readers, it reported a non-zero distance when comparing two identical 16-node graphs, as shown in Figure~\ref{fig:hunatcha_cfgs} in Appendix.

For more accurate CFG measurement, we extended the CFGED algorithm by Basque \textit{et al.}~\cite{basque2024ahoy} to handle addressless graphs.
The CFGED algorithm relies on a partial mapping of nodes across two CFGs, commonly collected by matching the addresses of node pairs.
After obfuscation and subsequent deobfuscation, this address association is lost.
Instead, we automatically created a mapping using information from the CFG blocks.
To do this, we extended the discovRE~\cite{eschweiler2016discovre} block-similarity algorithm.
First, we matched all nodes with the same function calls.
When multiple nodes match, we select pairs at the same distance from the graph's root.
After this initial set of mappings, we ran the discovRE algorithm to match any nodes more than 60\% similar.
We manually validated that all of these mappings were correct on all of our samples, and will release our tooling.
To further ensure correctness, we evaluated the effectiveness of our extended CFGED algorithm on CFGs used to evaluate CFGED in previous work~\cite{basque2024ahoy}.
This set consisted of 491 unique function pairs across optimized and unoptimized binaries from Coreutils.
Of the 491 pairs, our algorithm had the exact score on 389 pairs (79\%), while the Hu algorithm was exact on 36 pairs (7\%).
Additionally, our algorithm introduced an average similarity error of 3\% while the Hu algorithm introduced 22\%.

For CTF challenges lacking ground truth, we cannot perform graph edit distance matching.
Instead, we approximate similarity by collecting an API trace from a single execution and verifying whether our recovered CFG can produce the same API sequence through DFS traversal.

\smallskip
\noindent
\textbf{Results.}
Table~\ref{tab:merged-eval-alternative} shows the normalized similarity score produced by our extended CFGED algorithm.
We normalized our GED score as described in prior work~\cite{yadegari2015generic}.

\noindent\textit{Malware samples:} \system generated isomorphic CFGs for four of six samples, with near-isomorphic results for \textit{huffman} and \textit{netsky}.
The differences stem from redundant if-else branches that \system's simplification does not eliminate.

We evaluated the artifacts from prior work~\cite{yadegari2015generic} using our extended CFGED algorithm.
As Table~\ref{tab:merged-eval-alternative} shows, \system outperforms Yadegari's across all released artifacts.
The best similarity score Yadegari et al. achieved was 53\%, whereas \system consistently achieves significantly higher scores.

\noindent\textit{Open-source programs:} \system generated isomorphic graphs for both \textit{zlib} and \textit{sdl}. For \textit{curl}, the recovered CFGs closely match the original structure, with only minor simplifications remaining. The \textit{libpng} similarity score is lower due to \system's aggressive pruning of semantically empty nodes. Despite this, the recovered CFGs preserve overall structure and semantic correctness.

\begin{figure}[tb]
    \centering
    \includegraphics[width=0.73\linewidth]{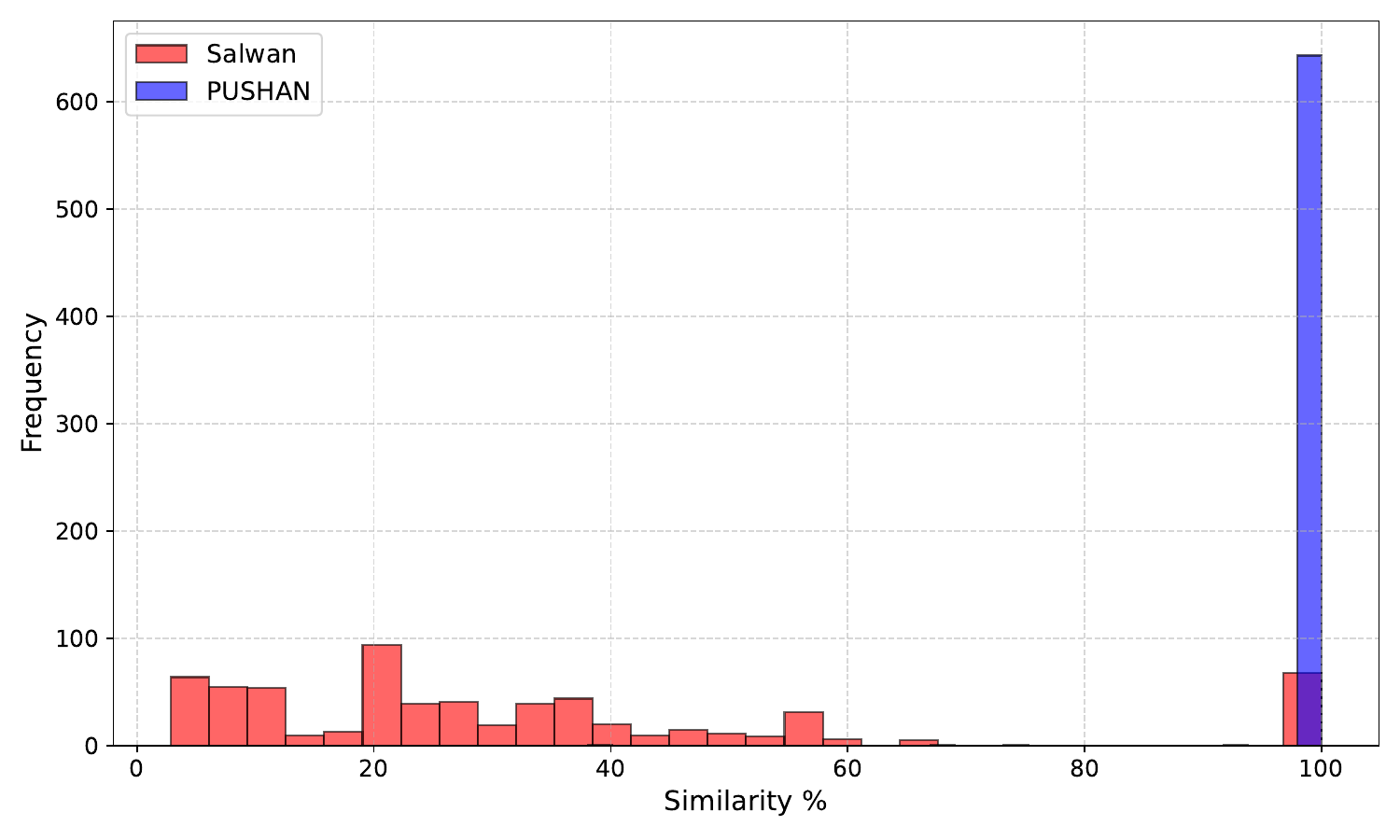}
    \caption{Histogram comparing CFG similarity produced by Salwan and PUSHAN over 647 common samples.}
    \label{fig:similarity-histogram}
\end{figure}

\noindent\textit{Tigress:} \system generated isomorphic graphs for 982 of 1,000 Tigress samples, as detailed in Table~\ref{tab:tigress-results}.
When evaluating Salwan's tool~\cite{salwan2018symbolic} on the same dataset, only 647 samples completed within three days.
As illustrated in Figure~\ref{fig:similarity-histogram}, when comparing CFG similarity on these 647 common samples, \system significantly outperforms Salwan’s tool---achieving 643 isomorphic graphs compared to only Salwan's 68.

Overall, \system restored isomorphic CFGs for 999 of 1,028 VM-obfuscated targets across all obfuscators.

\begin{table}[tb]
    \centering
    \caption{Tigress Deobfuscation and Decompilation Results across three VM configurations. LoC = lines of code.
    ``Verified'' indicates successful semantic validation.}
    \label{tab:tigress-results}
    {\footnotesize
    \begin{tabular}{lccc}
    \toprule
    \textbf{Metric} & \textbf{VM-1} & \textbf{VM-2} & \textbf{VM-3} \\
    \midrule
    \multicolumn{4}{l}{\textit{Deobfuscation Results}} \\
    Total Files               & 334         & 333         & 333         \\
    Succeeded                 & 330         & 329         & 329         \\
    Failed                    & 4           & 4          & 4           \\
    Success Rate              & 98.80\%     & 98.80\%     & 98.80\%     \\
    Total Time                & 1397.94 hr  & 2523.97 hr  & 1525.01 hr  \\
    Average Time              & 4.19 hr     & 7.58 hr     & 4.58 hr     \\
    \midrule
    \multicolumn{4}{l}{\textit{Verification Results}} \\
    Files Processed           & 330         & 329         & 328         \\
    Succeeded                 & 318         & 320         & 318         \\
    Failed                    & 12          & 9           & 10          \\
    Success Rate              & 96.36\%     & 97.26\%     & 96.95\%     \\
    Total Time                & 1430.73 hr  & 1335.43 hr  & 1717.03 hr  \\
    Average Time              & 4.34 hr     & 4.06 hr     & 5.23 hr     \\
    \midrule
    \multicolumn{4}{l}{\textit{Decompilation Quality}} \\
    Decompiled Files          & 330         & 329         & 329         \\
    Avg. LoC (Original)       & 187         & 185         & 188       \\
    Avg. LoC (Obfuscated)     & 4824        & 4686        & 5166      \\
    Avg. LoC (Decompiled)     & 692         & 739         & 698      \\
    Max Lines (Original)      & 555         & 604         & 516         \\
    Max Lines (Obfuscated)    & 13619       & 13246       & 12827       \\
    Max Lines (Decompiled)    & 2172        & 2506        & 2273        \\
    \midrule
    \multicolumn{4}{l}{\textit{CFG Similarity}} \\
    100\% Similarity (Count/Total)  & 328 / 330         & 326 / 329         & 328 / 329         \\
    \bottomrule
    \end{tabular}
    }
\end{table}

\subsection{Correctness: Semantic Equivalence}
\label{sec:eval_semcorrectness}

We evaluate \system's correctness using Tigress-generated programs because they compute hash functions, which provide a simple, deterministic mapping from inputs to outputs.
This enables automated validation through output comparison—if the deobfuscated program produces the same hash value as the original for a given input, behavioral equivalence is guaranteed.

We evaluated semantic correctness on Tigress samples using input/output testing on the recovered VEX IR.
Since Tigress programs compute hash functions, we tested each sample with 100 distinct inputs and compared the deobfuscated VEX IR outputs against the original binary.

We present our results in Table~\ref{tab:tigress-results}.
The recovered IR produced correct outputs in nearly all successful deobfuscation cases. Specifically, 956 of 1,000 Tigress samples passed all 100 input/output tests, demonstrating that \system reliably preserves functional behavior.
In some cases, decompilation failed while CFG-level verification succeeded.

\subsection{Analyzing A Real Binary via Code Reuse}
\label{sec:eval_codereuse}

We analyzed a VMProtect-obfuscated malware sample obtained in the wild.
According to VirusTotal, 28 out of 72 antivirus engines flagged this sample as malicious~\cite{diskgen_crack}, revealing a wide disagreement among them.
From reversing the non-obfuscated logic in this binary, we found that it uses a VM-obfuscated function to decrypt an embedded AutoIt script.
However, attackers may have also implemented malicious behaviors in the obfuscated function.
We must fully deobfuscate this function to determine whether the binary is malicious or not.

\system took a total of 85 minutes to deobfuscate and decompile the obfuscated function.
The final pseudo-code (476 lines) correctly reflected the complete control flow, revealing no additional malicious behaviors other than conducting integer operations inside two loops.

We provided the pseudocode output to Claude (Opus 4.6) with instructions to produce a simplified, recompilable, and semantically equivalent C implementation.
Notably, we gave the model only \system's decompiled pseudo-code, without access to the original binary, disassembly, or any auxiliary analysis tools.
Claude reduced the 476-line function to approximately 30 lines of recompilable C, inferring that the function is ``a pseudo-random number generator wrapper with hash-based seeding.''
This reduced function allowed us to extract the embedded AutoIt script.
We then realized that the main feature of this script was patching another commercial utility, DiskGenius, to potentially bypass its license checks if the host machine is not using Simplified or Traditional Chinese codepages.
This allows us to conclude that the binary is a crack, which is not inherently malicious but may be used for software piracy.

We note that while it is possible to extract the the embedded AutoIt script by debugging the binary and bypassing potential runtime protections (e.g., anti-debugging checks), dynamic analysis cannot reveal hidden behaviors in the VM-obfuscated function.
This highlights the necessity of static deobfuscation and decompilation of \system.

% We also evaluated whether Claude Opus 4.6 could simplify the function when given the original obfuscated binary directly, without \system's decompiled output.
% In this scenario, the model was provided with the binary and explicitly permitted to use any tools at its disposal, including disassemblers, decompilers, and other analysis utilities.
% While the model was able to produce a plausible simplification, it did so not by deobfuscating the virtualized code but by reasoning about the surrounding context in which the protected function was called.
% Although the result happened to be correct in this case, such an approach is inherently unreliable:
% A VM-protected function may implement additional malicious behavior, such as anti-analysis checks, data exfiltration, or environment fingerprinting, that is invisible from calling context alone.
% This highlights the necessity of \system's deobfuscation and decompilation:
% It enables analysis grounded in the actual semantics of the protected code rather than inference from external usage patterns.
% 
% This result demonstrates that \system successfully recovers the original program's control flow and semantics from the virtualized representation, and that the resulting output, despite containing residual obfuscation artifacts inherent to the protected binary, preserves sufficient semantic structure for state-of-the-art language models to recover high-level algorithmic intent and produce reusable source code.

\subsection{Extensibility: Decompiling Custom VMs}
\label{sec:eval_customvm}

Without ground-truth source code for CTF challenges, we validated correctness through API trace matching and comparison with public write-ups.
All five challenges matched both API traces and write-up descriptions (Table~\ref{tab:ctf-api-trace}).
In one challenge (Listing~\ref{lst:flag_check_compact}, appendix), the decompiled output directly revealed embedded flag values in conditional branches.

\noindent
\textit{CFG similarity.}
For all CTF challenges, we successfully reproduced the observed API call sequence by traversing the recovered CFG, confirming at least one correct execution path.
We also manually verified that each recovered CFG contained the functionality described in public write-ups.

\section{Discussion}
\label{s:disc}
%-------------------------------------------------------------------------------

\system does not yet simplify complex or obfuscated arithmetic expressions such as MBA expressions.
Without simplifying away obfuscated expressions, the decompilation output of \system may have a simplified control flow, but its semantics can still be hard to understand.
% This lack of MBA simplification can also lead to incorrect results, for certain analyses and also prevent further simplification of expressions.
Existing solutions (e.g., MBA-Blast) simplify many types of MBA expressions.
% From an engineering aspect, \system can integrate these solutions to further improve the deobfuscation and decompilation output.
LOKI~\cite{schloegel2022loki} primarily uses MBA expressions to obfuscate the semantics of handlers, without focusing on obfuscating the VM structure itself.
As \system does not support MBA simplification, we do not compare against these works.

Obfuscators can attack the symbolization process of \system by adding malformed control flows that result in the incorrect symbolization of actual constants.
This could cause \system to generate a deobfuscated CFG with missing branches or spurious branches that are not present in the original program.
We can improve the \system prototype by tracking the constant values more accurately:
Avoiding merging symbolic variables to \texttt{TOP} and instead introducing a value domain where values are guarded by their path predicates.
This way \system can split these merged values back to their original values and prevent incorrect symbolization.

Although \system is more scalable than deobfuscation solutions that rely on full-trace DSE, the current prototype of \system is still too slow due to the reliance on theorem provers to simplify MBA expressions to constants.
Integrating existing MBA simplification techniques would greatly speed up \system by avoiding excessive use of the solver. 
% \system requires accurate tracking of the locations of all VPCs.
% Although there are existing works that handle finding the virtual program counter ~\cite{sharif2009automatic}, it may be possible for obfsucators to make it harder to track the virtual program counter.

\section{Related Work}%
\label{s:related_work}

\noindent
\textbf{Virtualization deobfuscation.}
Kochberger~et~al.~\cite{kochberger2021sok} provide a systematization of various VM deobfuscation techniques based on extracted artifacts, analysis efforts, degree of automation, and generalizability. Beyond the works on automated deobfuscation studied in Section~\ref{sec:background:vmdeobfuscation}, Liang~et~al.~\cite{liang2018deobfuscation} propose to use symbolic execution to identify handler semantics. However, this comes with the pitfalls of using symbolic execution for deobfuscation that \pushan avoids. Similar to other works, SEEAD~\cite{tang2017seead} proposes dynamic taint analysis to identify control dependencies in obfuscated code, combined with aggressive path pruning. This increases performance but comes with the risk of missing paths.

\paragraph{MBA Deobfuscation}
Orthogonal to our approach, various techniques focus on the simplification of MBA expressions, including techniques such as: (a) Pattern matching~\cite{eyrolles2016defeating,biondi2017effectiveness}, which relies on existing knowledge, limiting their scope; 
(b)  Program synthesis~\cite{blazytko2017syntia,david2020qsynth,lee2023simplifying,menguy2021xyntia} generates simpler equivalent expressions but doesn't guarantee correctness;
%ProMBA3 combines synthesis and reweiting; 
(c) Algebraic methods~\cite{liu2021mbablast, reichenwallner2022simba, reichenwallner2023gamba} that are based on the property that two semantically equivalent n-bit input variables are aligned for 1-bit input variables, which is effective for linear MBA expressions;
(d) Deep learning~\cite{feng2020neureduce}, which takes a novel approach by training models to deobfuscate MBA expressions.
Additionally, Arybo~\cite{guinet2016arybo} employs the Bit-Blast method, simplifying arithmetic expressions to bit-level symbolic expressions, best suited for small expressions because of high-performance cost.
% These techniques can be used with \system to simplify MBA expressions.

%-------------------------------------------------------------------------------
\section{Conclusion}
\label{s:conclusion}
%-------------------------------------------------------------------------------

\system is a scalable technique for deobfuscating and decompiling virtualization-obfuscated binaries.
Building on the concept of VPC-sensitive CFG recovery, \system allows for the recovery of complete, pre-obfuscation CFGs without relying on user input that covers all branches of the obfuscated program or performing DSE on a sufficient number of traces and finding user input.
The experiments show that \system can recover the CFGs of the original, unobfuscated programs with a high level of similarity and output decompiled code that can enable advanced malware analysis with LLMs.
% We hope \system will benefit the reverse engineering of virtualization-protected malware samples.


% \section*{Acknowledgment}



\section*{Ethical Considerations}

% This work aims to advance defensive security capabilities through improved malware analysis. We acknowledge the dual-use nature of deobfuscation research: while our techniques could theoretically help malicious actors understand protection mechanisms, we believe the benefits to defenders significantly outweigh potential harms. Virtualization-based obfuscation is widely used for legitimate purposes such as protecting intellectual property and preventing game piracy, but is also heavily employed by malware authors. Our work primarily benefits security analysts who need to analyze protected malware at scale, as manual reverse engineering is time-consuming, error-prone, and does not scale to the volume of obfuscated malware encountered in practice.

% We conducted all experiments on publicly available datasets and malware samples from public sources.
% Our analysis of the VMProtect-obfuscated sample revealed it to be a software crack rather than intentional malware.
% No human subjects or sensitive data were involved in our research.
% We conclude that publication of our research benefits the security community by enabling more effective analysis of protected malware, which is essential for developing defenses against evolving threats.

This work aims to advance defensive security capabilities through improved malware analysis. 

\paragraph{Stakeholders} We identify three stakeholders, malware analysts, companies relying on virtualization-obfuscation for legitimate purposes, and society in general.
% Security analysts are our primary target group and benefit from our technique, as it boosts their effectiveness towards analyzing malicious code. 
% At the same time, companies also use virtualization to protect their intellectual property, address cheating in online games, or use it for Digital Rights Management (DRM). Thus, these companies are our second stakeholders. 
% Our third stakeholder is society at large, which will profit of better analysis of malware and timely mitigation.

\paragraph{Impact} Our work has both a positive and negative impact.

\noindent\emph{Positive Impact:} Our work helps malware analysts to analyze protected malware at scale, as manual reverse engineering is time-consuming, error-prone, and does not scale to the volume of obfuscated malware encountered in practice. Helping to identify malware enables timely mitigation, thus benefitting society in general. 

\noindent\emph{Negative Impact:} Malicious actors could likewise use our code to target benign code obfuscated by companies to protect their intellectual property, prevent cheating in online games, or use it for Digital Rights Management (DRM) applications.

% Following the Menlo report, we consider beneficence, respect for persons, justice, and respect for law and public interest as ethical principles. We believe our work to be of significant help for malware analysts, leading to long-term benign consequences for society in general (beneficence). These positives benefits outweigh the negative risk for companies employing virtualization as protection. Our research respects persons, as it does not involve or target individual actors and we do not process any personal data. Justice  

\paragraph{Mitigations} 
To avoid harm, we conducted all experiments on publicly available datasets and malware samples from public sources.
Our analysis of the VMProtect-obfuscated sample revealed it to be a software crack rather than intentional malware.
No human subjects or sensitive data were involved in our research, ensuring we respect persons.
Ultimately, we cannot prevent misuse of our tool by malicious actors.
% In terms of misuse of our tool, there is no practical mitigation: After release, any actor may use our tool, including for malicious purposes. However, as manual deobfuscation of VM-based obfuscation is already widespread practice, our tool helps the actors needing to work with such software at scale: malware analysts.

\paragraph{Justification for Research} 
We believe the benefits to defenders significantly outweigh potential harms. 
Virtualization-based obfuscation is widely used for legitimate purposes such as protecting intellectual property and preventing game piracy, but is also heavily employed by malware authors. 
Our work primarily benefits security analysts who need to analyze protected malware at scale, as manual reverse engineering is time-consuming, error-prone, and does not scale to the volume of obfuscated malware encountered in practice. 
We believe that publication of our research benefits the security community by enabling more effective analysis of protected malware, which is essential for developing defenses against evolving threats.

\section*{Acknowledgments}

This work was supported by the Advanced Research Projects Agency for Health (ARPA-H) under Contract No. SP4701-23-C-0074, the National Science Foundation (NSF) under Grants No. 2232915 and 2146568, and the Office of Naval Research (ONR) under Grant No. N00014-23-1-2563. We also gratefully acknowledge the generous support of the U.S. Department of Defense.

\bibliographystyle{plainurl}
\bibliography{strings,references}

\appendix
\section*{A. The Original and Deobfuscated CFGs of Some Samples}
\label{sec:cfg_comparison}

\begin{figure*}[htbp]
    \centering
    \begin{subfigure}[b]{0.80\linewidth}
        \centering
	     \rotatebox{90}{% Rotate the entire content of the subfigure
             \begin{minipage}{0.55\textwidth}
	    \includegraphics[width=\linewidth]{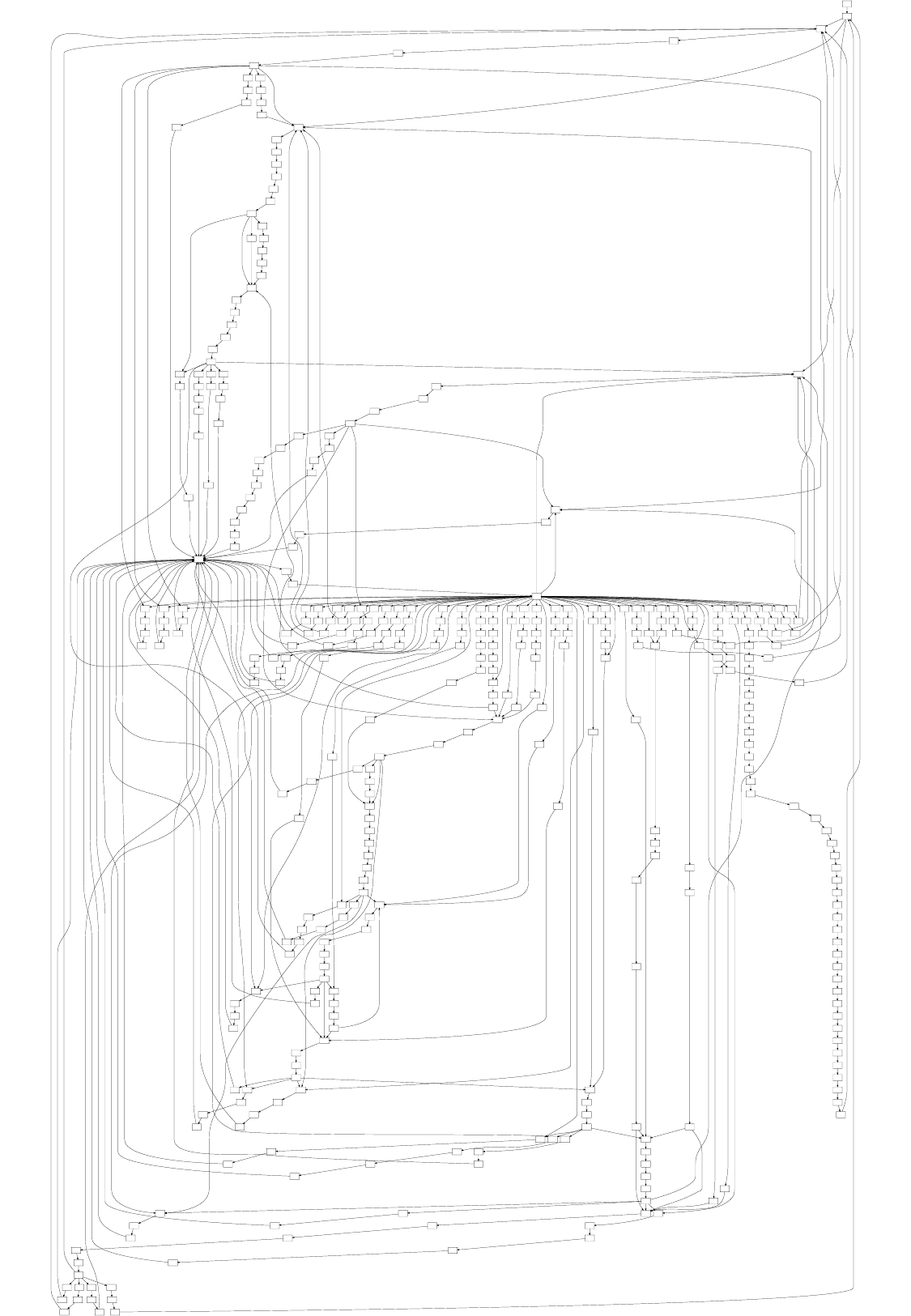}
        \caption{The CFG of the VMProtect-obfuscated huffman program. Cropped for clarity.}
	\end{minipage}
            }

    \end{subfigure}
    % No space command here
    \begin{subfigure}[b]{.56\linewidth}
        \centering
	    \rotatebox{90}{% Rotate the entire content of the subfigure
	     \begin{minipage}{0.65\textwidth}
	    \includegraphics[width=\linewidth]{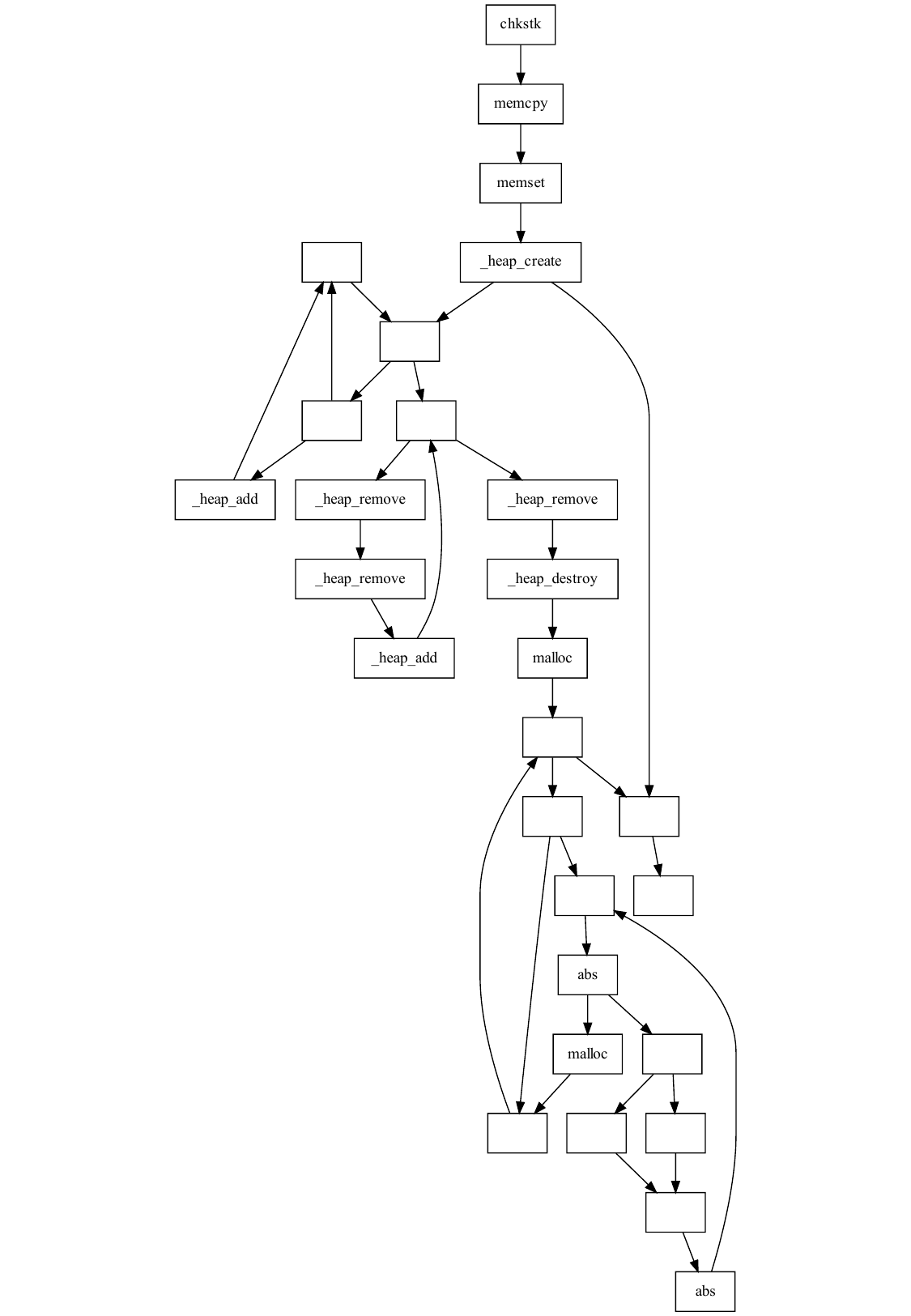}
	    \caption{The CFG of the original huffman program.}
		\end{minipage}
	    }
    \end{subfigure}
    % Adding the third subfigure
    \begin{subfigure}[b]{0.56\linewidth}
        \centering
	     \rotatebox{90}{% Rotate the entire content of the subfigure
             \begin{minipage}{0.65\textwidth}

	    \includegraphics[width=\linewidth]{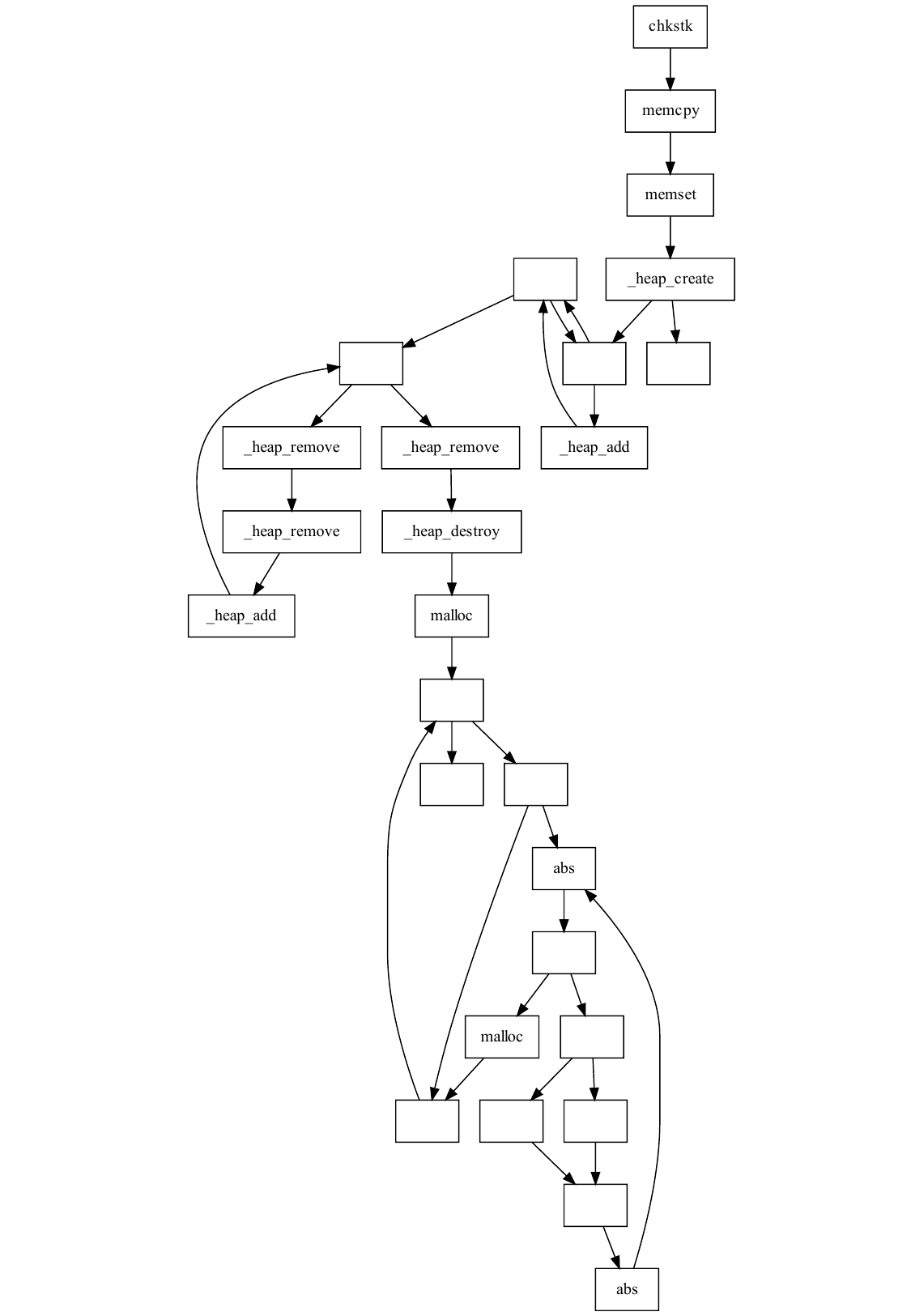}
        \caption{The deobfuscated CFG of the huffman program.}
\end{minipage}
            }

    \end{subfigure}
	\caption{A comparison of the original, obfuscated, and deobfuscated CFGs for huffman.}
    \label{fig:huffman_cfgs}
\end{figure*}

\begin{figure*}[htbp]
    \centering
    \begin{subfigure}[b]{0.80\linewidth}
        \centering
	     \rotatebox{90}{% Rotate the entire content of the subfigure
             \begin{minipage}{0.55\textwidth}

        \includegraphics[width=\linewidth]{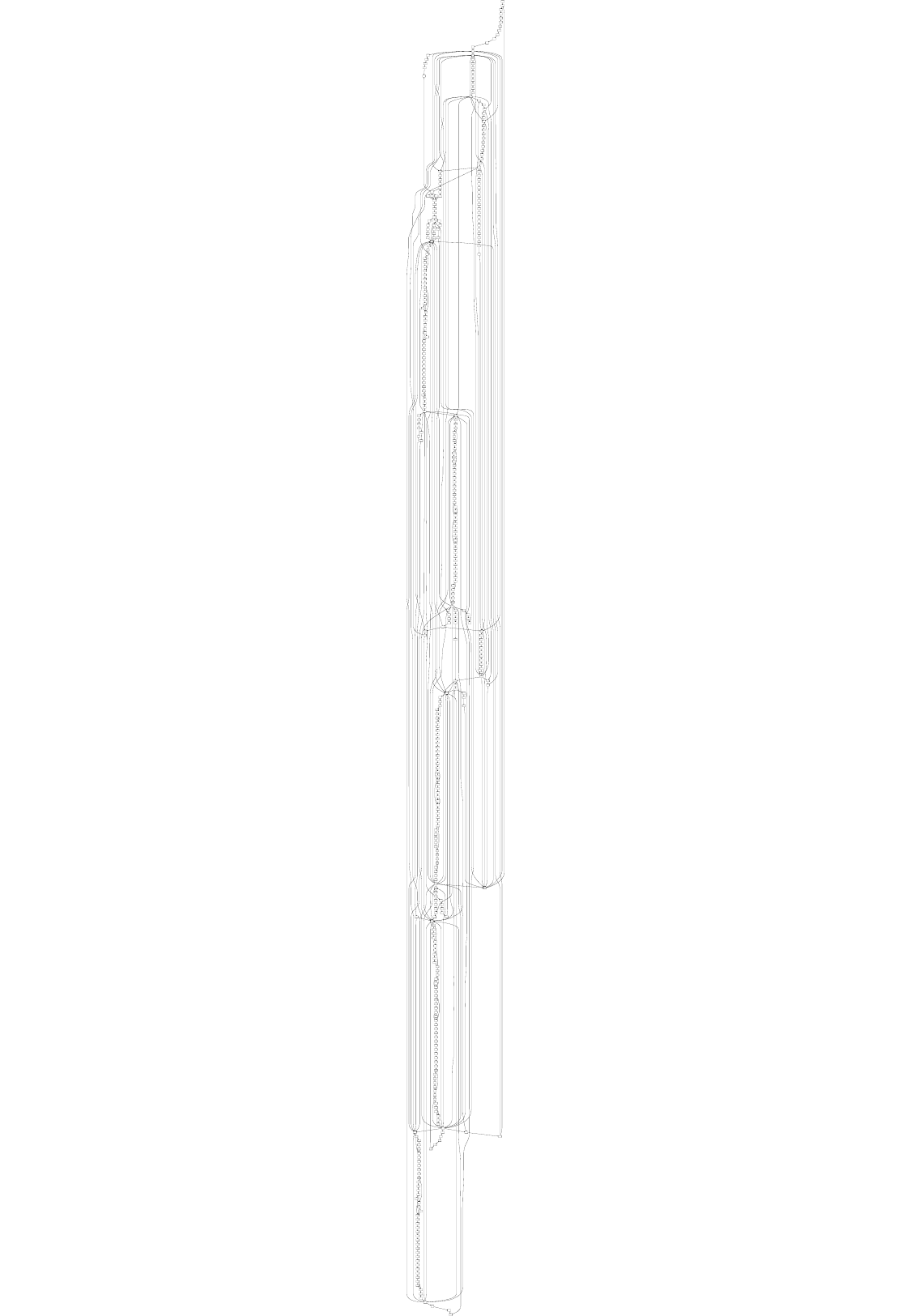}
        \caption{The CFG of the Themida-obfuscated Hunatcha program.}
		            \end{minipage}
            }
    \end{subfigure}
    % No space command here
    \begin{subfigure}[b]{0.56\linewidth}
        \centering
	    \rotatebox{90}{% Rotate the entire content of the subfigure
             \begin{minipage}{0.65\textwidth}

        \includegraphics[width=\linewidth]{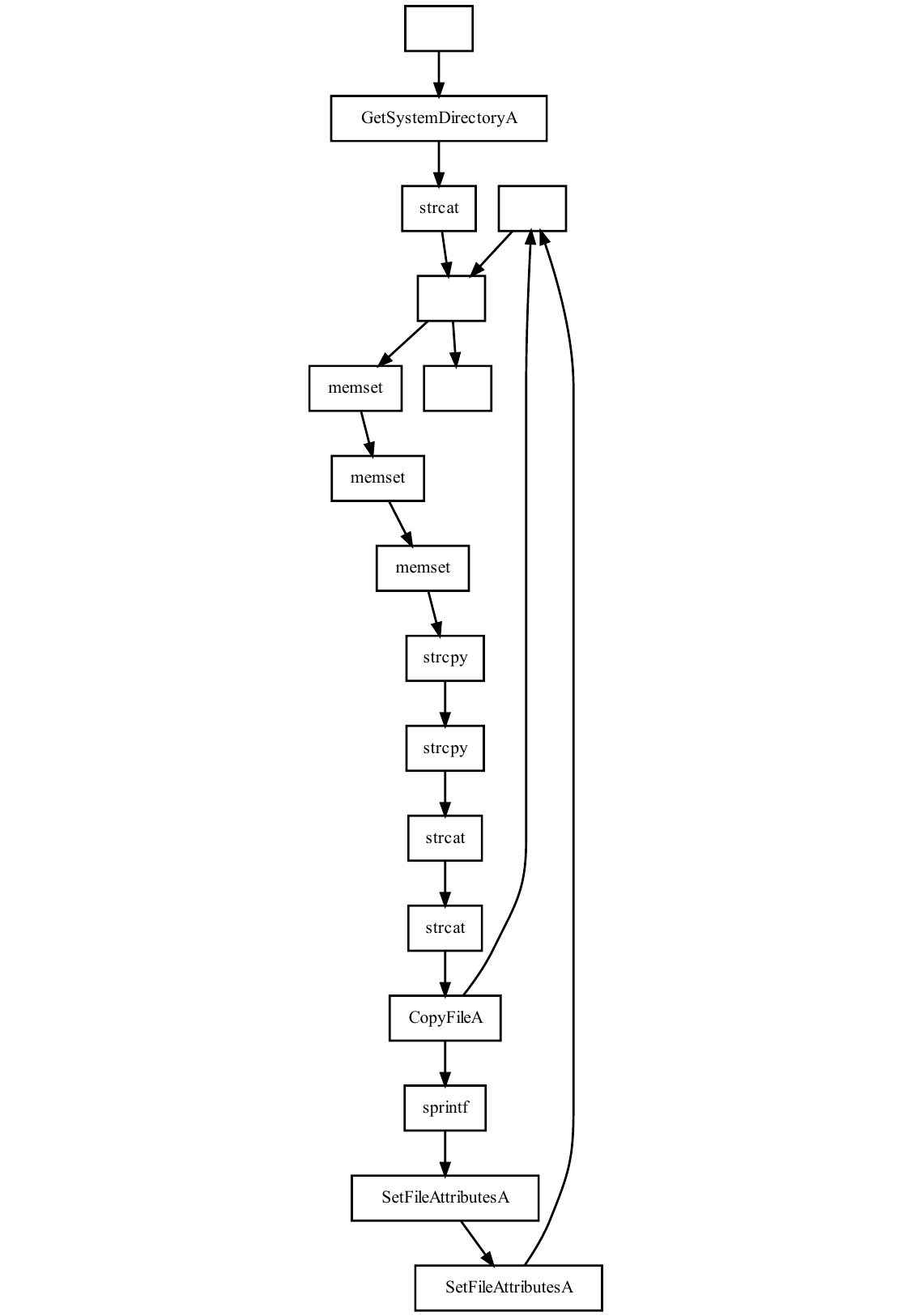}
        \caption{The CFG of the original Hunatcha program.}
		                         \end{minipage}
            }

    \end{subfigure}
    % Adding the third subfigure
    \begin{subfigure}[b]{0.56\linewidth}
        \centering
	    \rotatebox{90}{% Rotate the entire content of the subfigure
             \begin{minipage}{0.65\textwidth}

        \includegraphics[width=\linewidth]{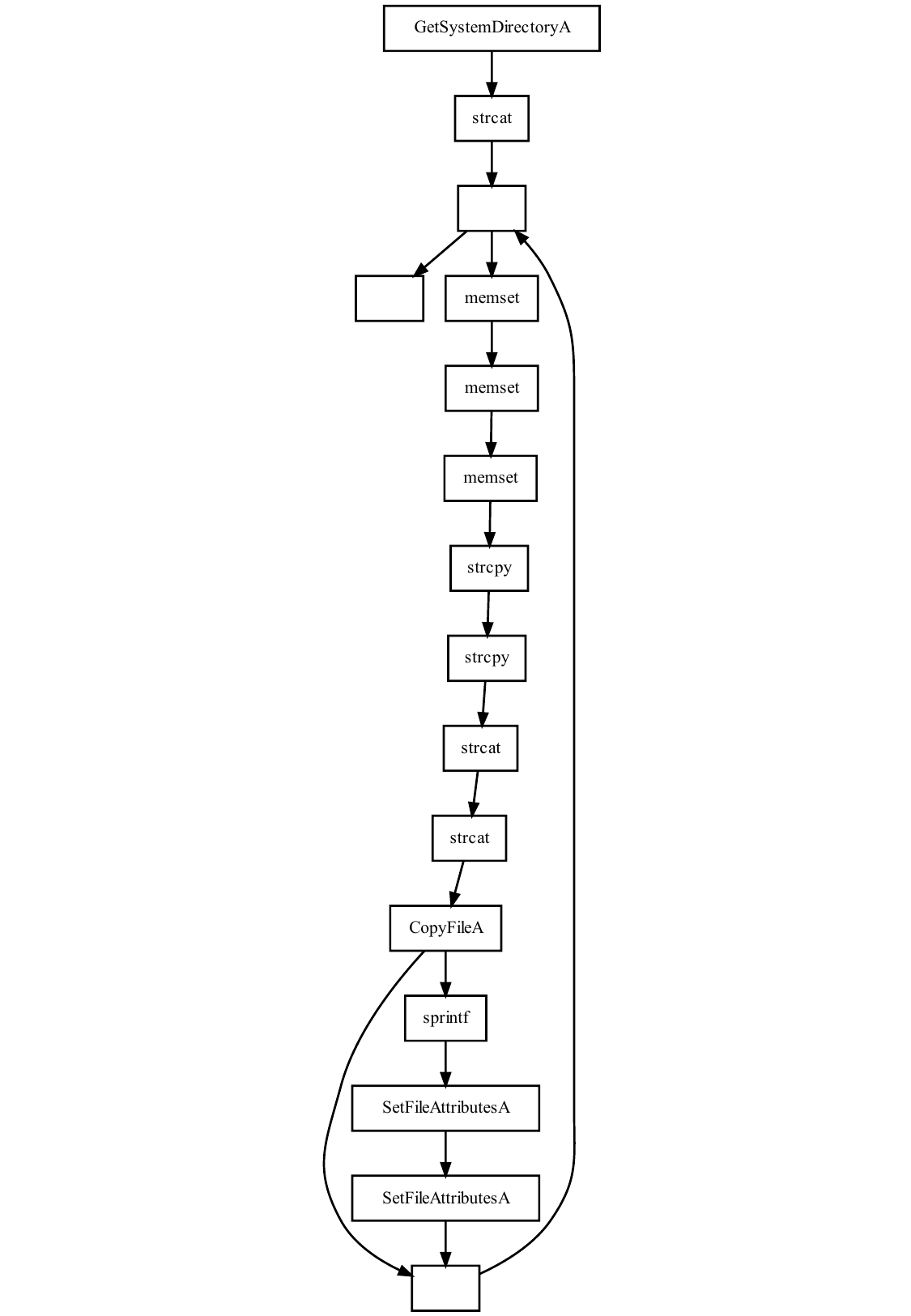}
        \caption{The deobfuscated CFG of the original Hunatcha program.}
		                         \end{minipage}
            }

    \end{subfigure}
    \caption{A comparison of the original, obfuscated, and deobfuscated CFGs of hunatcha.}
    \label{fig:hunatcha_cfgs}
\end{figure*}

Figures~\ref{fig:huffman_cfgs} and \ref{fig:hunatcha_cfgs} provide a visual comparison between the original and the deobfuscated CFGs of three samples.

\begin{figure*}[tbp]
    \centering
    \begin{subfigure}[b]{0.45\linewidth}
        \centering
        \begin{lstlisting}[basicstyle=\ttfamily\scriptsize, aboveskip=0pt, belowskip=0pt, xleftmargin=15pt]
t2 = GET:I32(eax)                   
t0 = Add32(t2, 0x00000001)        
t3 = Add32(t0, 0x00000002)
        \end{lstlisting}
        \caption{Before arithmetic simplification.}
    \end{subfigure}
    % Space can be adjusted as needed using \hspace or similar commands
    \hfill
    \begin{subfigure}[b]{0.45\linewidth}
        \centering
        \begin{lstlisting}[basicstyle=\ttfamily\scriptsize, aboveskip=0pt, belowskip=0pt, xrightmargin=5pt]
t2 = GET:I32(eax)

t3 = Add32(t2, 0x00000003)
        \end{lstlisting}
        \caption{After arithmetic simplification.}
    \end{subfigure}
    
    \vspace{0.5em} % Adding vertical space between rows of subfigures
    
    \begin{subfigure}[b]{0.45\linewidth}
        \centering
        \begin{lstlisting}[basicstyle=\ttfamily\scriptsize, aboveskip=0pt, belowskip=0pt, xleftmargin=15pt]
push addr
ret
        \end{lstlisting}
        \caption{Before \code{push retn} simplification.}
    \end{subfigure}
    \hfill
    \begin{subfigure}[b]{0.45\linewidth}
        \centering
        \begin{lstlisting}[basicstyle=\ttfamily\scriptsize, aboveskip=0pt, belowskip=0pt, xrightmargin=5pt]
jmp addr
        \end{lstlisting}
        \caption{After \code{push retn} simplification.}
    \end{subfigure}

    \vspace{0.5em}
    
    \begin{subfigure}[b]{0.45\linewidth}
        \centering
        \begin{lstlisting}[basicstyle=\ttfamily\scriptsize, aboveskip=0pt, belowskip=0pt, xleftmargin=15pt]
t29 = Add64(t12,0xfffffffffffffe80)
STle(t29) = t27
...
t31 = Add64(t23,0xfffffffffffffe80)
t33 = LDle:I64(t31)        
        \end{lstlisting}
        \caption{Redundant stack-based data movement.}
    \end{subfigure}
    \hfill
    \begin{subfigure}[b]{0.45\linewidth}
        \centering
        \begin{lstlisting}[basicstyle=\ttfamily\scriptsize, aboveskip=0pt, belowskip=0pt, xrightmargin=5pt]
t29 = Add64(t12,0xfffffffffffffe80)
STle(t29) = t27
...
t31 = Add64(t23,0xfffffffffffffe80)
t33 = t27
        \end{lstlisting}
        \caption{After eliminating redundant stack-based data movement.}
    \end{subfigure}

    \caption{Examples of semantics-preserving simplifications.}%
    \label{fig:semantic_preserving_code_simplifications} % Label for referencing
\end{figure*}

\section*{B. CTF Challenge}

\begin{listing}[tb]
    \centering
\begin{lstlisting}[
        language=C,
        basicstyle=\ttfamily\scriptsize,
        numbers=left,
        numbersep=3pt,
        frame=lines,
        breaklines=true,
        breakindent=1em
    ]
cmp = 1885566054 - input;  // "pctf"
if (!cmp) return;
// ... omitted code ...
cmp = 2071358815 - input;  // "vm_{"
if (!cmp) return;
// ... omitted code ...
cmp = 1915975269 - input;  // "r3ve"
if (!cmp) return;
// ... omitted code ...
cmp = 1920152425 - input;  // "rs3i"
if (!cmp) return;
// ... omitted code ...
cmp = 1850171185 - input;  // "nG_1"
if (!cmp) return;
// ... omitted code ...
cmp = 1935635570 - input;  // "s_tr"
if (!cmp) return;
// ... omitted code ...
cmp = 828599161 - input;   // "1cky"
if (!cmp) return;
// ... omitted code ...
cmp = 2100310064 - input;  // "}000"
if (!cmp) return;
\end{lstlisting}
    \vspace{-0.5em}
    \caption{
    Decompiled output from a CTF challenge implementing a custom VM.
    Each constant maps to a 4-character segment of the flag, which can be directly recovered by inspecting the arithmetic conditions.}
    \label{lst:flag_check_compact}
\end{listing}

\section*{C. \system Deobfuscation Output of The Real-world Sample}

\begin{listing}[tb]
    \centering
\begin{lstlisting}[
        language=C,
        basicstyle=\ttfamily\tiny,
        numbers=left,
        numbersep=3pt,
        frame=lines,
        breaklines=true,
        breakindent=1em,
        firstline=1,
        lastline=100
    ]
extern char g_7ff74db5f1e5;
extern char g_7ff74db5f705;
extern char g_7ff74db5ff1a;

long VM_1(unsigned long a0, unsigned int a1)
{   
    .
    .
    .
    .

    v44 = 0;
    v43 = v25;
    v7 = *((int *)&v43);
    v8 = 0;
    v42 = 8 + v9;
    v19 = v42;
    v15 = v19;
    v24 = 0x7ff60da70000;
    v6 = tmp_1035;
    v5 = v9;
    v9 = v19;
    v19 = 17;
    v2 = *((long long *)&v7);
    v23 = v10;
    v7 = v15;
    v17 = tmp_1070;
    v10 = v22;
    v47 = 435308053;    
    .
    .
    .
    .
    v19 = v36;
    v36 = v24 + 68;
    v13 = v36;
    v14 = 0;
    memmove(v19, v18, *((long long *)&v13));
    v13 = v20;
    v19 = v18;
    v18 = 0;
    v6 = 144 + v13;
    v91 = 1;
    v92 = (v18 >> 21 == 1 ? 1 : 0);
    v93 = (v18 >> 18 == 1 ? 1 : 0);
    memmove(v6, v19, 68);
    v5 = 0;
    v40 = v13;
    v33 = v5;
    v18 = 9;
    do
    {
        v50 = 4 | v91 & 0x400 | v92 * 0x200000 & 0x200000 | v93 * 0x40000 & 0x40000;
        v35 = v18;
        v34 = v33;
        v91 = (v34 >> 10 == 1 ? -1 : 1);
        v92 = (v34 >> 21 == 1 ? 1 : 0);
        v93 = (v34 >> 18 == 1 ? 1 : 0);
        v94 = prng(v40);
        v21 = 0;
        v12 = v94;
        v18 = v40;
        v19 = 0;
        v20 = -1 + v35;
        v24 = (1 & v21) + (-2 & v19);
        v21 = ~(~(v24) & -17) & ~(16 & ~(~(v24) & ~((1 & v21) + (-2 & v19))));
        v50 = *((long long *)((char *)&<0x4e6f6e6531343037303031333639333939393230[is_189]|Stack bp-16, 1 B> + ((~(v21) & 64) >> 3)));
        v24 = v50;
        v49 = v50;
        v49 = ~(*((long long *)&v49));
        v50 = *((int *)((char *)&v49 + 4)) & (int)v49;
        v49 = 247274402;
        v49 = ~(*((long long *)&v49));
        v50 = *((int *)((char *)&v49 + 4)) & -247274403;
        v49 = -247274403;
        v48 = v24;
        v48 = ~(*((long long *)&v48));
        v49 = 247274402 & (int)v48;
        v49 = ~(*((long long *)&v49));
        v50 = *((int *)((char *)&v49 + 4)) & (int)v49;
        v13 = v91 & 0x400 | v92 * 0x200000 & 0x200000 | v93 * 0x40000 & 0x40000;
        v33 = v21;
        v32 = v13;
        v27 = v50;
        v95 = (int)v27 + 0x100000000;
        v96 = *((char *)((int)v27 + 140699062697985));
        .
        .
        .
        .

    } while ((((((((&g_7ff74db5ff1a)[__ROR__(__ROL__(*((char *)((int)v27 + 140699062698017)) - ((char)v0 - 133 - (char)v102), 5) - 33, 4)] & -0xff00ff00ff0100) >> 8 | (&g_7ff74db5ff1a)[__ROR__(__ROL__(*((char *)((int)v27 + 140699062698017)) - ((char)v0 - 133 - (char)v102), 5) - 33, 4)] << 8 & -0xff00ff00ff0100) & -0xffff00010000) >> 16 | 0x400e000000000000) & -0x100000000) >> 32 | 0x100000000) + 0x7ff60da70000 == 140700137420549);
    v50 = 0;
    *((unsigned long *)&v50[88]) = v12;
    return (long long)v50[88];
}
\end{lstlisting}
    \vspace{-0.5em}
    \caption{
      \system's deobfuscated output (truncated) for the real-world sample~\cite{diskgen_crack}.}
    \label{lst:pushan_output}
\end{listing}

\section*{D. Claude's Simplified Output of the Real-world Sample}

\begin{listing}[tb]
    \centering
\begin{lstlisting}[
        language=C,
        basicstyle=\ttfamily\scriptsize,
        numbers=left,
        numbersep=3pt,
        frame=lines,
        breaklines=true,
        breakindent=1em
    ]
#include <string.h>
#include <stdint.h>

// External PRNG function (provided elsewhere in the binary)
extern uint64_t prng(uint64_t state);

long VM_1(unsigned long a0, unsigned int a1) {
    uint8_t *buf = (uint8_t *)a0;
    uint32_t count = a1;

    // Phase 1: Fill buffer with multiplicative hash sequence
    uint32_t *out = (uint32_t *)buf;
    uint32_t val = count;
    for (int i = 0; i < 17; i++) {
        val = 1403630842u * val - 2;
        // The NOT/AND pattern collapses to: val = val (with potential truncation)
        // effectively: val = (uint32_t)(~(uint64_t)val) & high32 => just keeps val
        val = val; // obfuscated identity after simplification
        *out++ = val;
    }

    // Phase 2: memmove 68 bytes within the buffer
    memmove(buf + 76, buf, 68);
    memmove(buf + 144, buf, 68);

    // Phase 3: PRNG loop
    uint64_t state = (uint64_t)(uintptr_t)buf;
    uint64_t result = 0;
    for (int i = 9; i > 0; i--) {
        result = prng(state);
    }

    // Store and return final PRNG result
    *(uint64_t *)(buf + 88) = result;
    return (long)result;
}
\end{lstlisting}
    \vspace{-0.5em}
    \caption{
    Claude's simplified version of \system's output for the real-world sample.}
    \label{lst:claude_simplified}
\end{listing}

\end{document}